\documentclass[pra,twocolumn,showpacs,preprintnumbers,amsmath,amssymb]{revtex4}
\DeclareMathOperator{\tint}{\textstyle\int}
\usepackage{bm,graphics}
\DeclareSymbolFontAlphabet{\mathbfcal}{boldsymbols}
\begin{document}
\title{Many-body theory of surface-enhanced Raman scattering}
\author{David J. Masiello}
\email{masiello@chem.northwestern.edu}
\author{George C. Schatz}
\email{schatz@chem.northwestern.edu}
\affiliation{Department of Chemistry, Northwestern University, Evanston, Illinois 60208-3113, USA}
%\date{\today} 
\date{June 11, 2008}   
\begin{abstract}
A many-body Green's function approach to the microscopic theory of surface-enhanced Raman scattering is presented. Interaction effects between a general molecular system and a spatially anisotropic metal particle supporting plasmon excitations in the presence of an external radiation field are systematically included through many-body perturbation theory. Reduction of the exact effects of molecular-electronic correlation to the level of Hartree-Fock mean-field theory is made for practical initial implementation, while description of collective oscillations of conduction electrons in the metal is reduced to that of a classical plasma density; extension of the former to a Kohn-Sham density-functional or second-order M{\o}ller-Plesset perturbation theory is discussed; further specialization of the latter to the random-phase approximation allows for several salient features of the formalism to be highlighted without need for numerical computation. Scattering and linear-response properties of the coupled system subjected to an external perturbing electric field in the electric-dipole interaction approximation are investigated. Both damping and finite-lifetime effects of molecular-electronic excitations as well as the characteristic fourth-power enhancement of the molecular Raman scattering intensity are elucidated from first principles. It is demonstrated that the presented theory reduces to previous models of surface-enhanced Raman scattering and leads naturally to a semiclassical picture of the response of a quantum-mechanical molecular system interacting with a spatially anisotropic classical metal particle with electronic polarization approximated by a discretized collection of electric dipoles.
\end{abstract}
\maketitle
\section{Introduction}
Inherently exceedingly weak, Raman scattering \cite{Raman1928} of incident electromagnetic radiation from a molecular target occurs, approximately, only once out of every million photon-molecule scattering events. These few inelastically scattered photons carry away a fraction of energy less or more than they had originally with the difference being deposited into or liberated from molecular vibrational, or, to a lesser extent rotational, excitation. Resultant changes in electronic polarizability with respect to underlying nuclear geometry ultimately are encoded in spectral fingerprints of molecular structure. Discerning such small Raman signals from a large elastic Rayleigh-scattering background is challenging, yet potentially rewarding, as Raman spectra can reveal detailed molecular structural information unresolved by other spectroscopies.

In the 1970s it was first observed \cite{Fleischmann74} and later understood \cite{vanduyne77,Creighton1977} that certain molecules, if adsorbed onto roughened noble-metal substrates having structure on the subwavelength scale, experience a surface-enhanced Raman scattering (SERS) of incident photons that manifests itself in approximately a millionfold boost in signal in comparison to the normal Raman effect. Out of this new phenomenon, the field of surface-enhanced Raman spectroscopy was rapidly born; early reviews can be found in Refs. \cite{Moskovits1985,Metiu1984,Kerker1984,Schatz1984b}. Two mechanisms are generally believed to account for enhancement of the Raman-scattered field: one of chemical and the other of electromagnetic origin. The former rests upon the idea that a chemical bond is formed between the adsorbed molecule and the metal, thus allowing for charge-transfer excitations to occur between the two systems. The latter, which is widely regarded as being dominant \cite{Moskovits2005}, involves the coupling of external radiation to surface-plasmon excitations at the metal-dielectric interface; plasmons are quantized collective oscillations of conduction electrons against the positive ionic background that can act to enhance and focus incident light to subwavelength dimension below the diffraction limit. When optically excited, these plasmons broadcast their enhanced field to nearby Raman-active molecules which inelastically Raman scatter photons back to the metal. The scattered photons can recouple into the plasmon modes of the metal and, subsequently, be rebroadcasted toward a detector. Both normal and surface-enhanced Raman scattering events are linear processes (depending only upon the linear polarizability) yet, if both incident and Raman scattered frequencies are resonant with plasmon excitations in the metal, their enhancements can multiply together to yield a fourth-power enhancement of the normal Raman-scattering intensity.

Recently, surface-enhanced Raman spectroscopy has gone through a renaissance which is, in part, attributable to the first observation of single-molecule SERS in the late 1990s \cite{Nie1997,Kneipp1997}. Exhibiting a giant boost in Raman signal (by 10 or more orders of magnitude in comparison to the normal Raman scattering from a single molecule in free space), single-molecule surface-enhanced Raman spectroscopy provides an even richer variety of molecular structural information that is free from ensemble averaging. With the ability to measure single-molecule Raman spectra, basic research has continued to progress under the impetus of utilizing SERS techniques, {\it inter alia}, as an ultra sensitive analytical probe having broad utility in the biological sciences; see, {\it e.g.}, Refs. \cite{Mirkin2002,Xu1999,Kneipp1998}.

Today, over thirty years after its initial discovery and a decade after the observation of single-molecule SERS, significant work is underway to systematically characterize the best conditions for single-molecule SERS activity in individual and arrays of nanoscale metal particles as a function of size, shape, and interparticle separation, among others, with respect to the wavelength of light \cite{Masiello08b}. Independent variation of each of these nanoscale characteristics is controllable in the laboratory, with the number of permutations exceedingly large. Guidance from predictive theory would be of immense utility in this pursuit, however, current theoretical methods cannot completely meet this challenge.

Involving the coupling and interaction of a Raman-active molecular system with one or more nanoscale metal particles under the influence of an external radiation field, the SERS effect presents a complicated many-body problem blending together concepts from quantum chemistry and molecular spectroscopy, condensed-matter physics, and electromagnetism. Undoubtedly, its present incomplete theoretical description is rooted in the complexity of its basic processes. As it not our intent to exhaustively summarize thirty years of theoretical progress, we defer to the reviews \cite{Moskovits2005,Moskovits1985,Metiu1984,Kerker1984,Schatz1984b} and briefly discuss only a few recent and notable approaches based upon the complementary starting points of classical and quantum mechanics \cite{Schatz2007a}. Almost all previous approaches can be placed into one of these two categories.

The optical and plasmonic properties of spatially anisotropic nanoscale metal particles (and particle arrays), which may have dimensions from tens to hundreds of nanometers, are well described within classical electromagnetic theory. Useful physical information can be gleaned from electrostatic model calculations \cite{Mie1908,Kerker1980} as well as from full numerical solution of Maxwell's equations \cite{Yee1966,Taflove}, such as the magnitude and location of enhanced electromagnetic fields located near the particle's surface \cite{Schatz2004a}: so called electromagnetic hot spots. However, while a classical description based upon the metal's underlying continuum dielectric function may be appropriate for a nanoscale particle, an adsorbed molecular system undergoing inelastic Raman scattering of photons is properly described only from a microscopic point of view.

To this end, recently, a fully quantum description of the combined nanoparticle-molecule system has been developed within the Kohn-Sham framework of time-dependent density-functional theory \cite{Jensen2006a}. Both particle and molecule are represented by the same total wave function. Leaving aside shortcomings inherent in the choice of electronic-correlation functional and its nonsystematic improvability, this approach is limited, due to computational restrictions, to the treatment of small metal particles containing, at most, on the order of one hundred atoms \cite{Aikens2008}. With such small numbers, metallic particles display the discrete electronic-excitation structure more typical of clusters than of the bulk-like resonance continuum exhibited on the nanoscale. In this single-particle picture, molecular electronically-excited states have an infinite lifetime as there is no mechanism for their damping due to the presence of the metallic system; this is a consequence of the fact that there is no explicit treatment of the interaction between molecule and particle beyond that specified in the single-particle Kohn-Sham formalism. {\it Ad hoc} empirical parametrization is employed to mimic these basic interactions, and, in turn, damp molecular-electronic excitation. In this way, deficiencies in the theory are corrected, leading to predictions which compare sensibly with experiment \cite{Jensen2006a,Aikens2006a,Jensen2007b}. Two additional notable quantum approaches to SERS based upon density-matrix calculations have recently appeared in the literature \cite{Johansson2005,Kelley2008}. The interaction of a Raman-active molecule supporting two electronic states (plus several vibrational substates) with two nearby nanoscale Ag spheres, described through an extended Mie theory, is presented in Ref. \cite{Johansson2005}. With appropriate choice of parameters and inclusion of phenomenological damping mechanisms, Raman-scattering cross sections enhanced by 10 orders of magnitude are demonstrated in comparison to the normal Raman effect. Second, in Ref. \cite{Kelley2008}, enhanced resonance Raman phenomena are studied within a combined  eight-state density-matrix approach where the molecular subsystem is represented in a four-state basis involving molecular ground and electronic, vibrational, and electronic and vibrational excited states while the particle subsystem is represented in a two-state basis consisting of ground and excited states. Within this model, it is predicted that the largest resonance Raman enhancements occur when a molecule, which absorbs light far from the particle's resonance maximum, is excited at the resonance maximum of the particle. This prediction is believed to be caused by the shifting of molecular resonances due to the strong coupling between molecule and particle.

Each of the approaches described above have both positive and negative attributes. It is therefore natural to envision blending their best features and, simultaneously, to explicitly treat the coupling between molecule and particle so as to avoid parametrization. It is the purpose of this article to do exactly that.

Here we present a formal {\it ab initio} many-body theory underlying a unified and didactic approach to the description of single-molecule SERS from a nanoscale metal particle at zero temperature. Emphasis is placed on developing a rigorous, yet computationally tractable formalism. In anticipation of practical initial numerical implementation, specialization is made, within the Born-Oppenheimer approximation, to a Hartree-Fock (HF) mean-field description of the electronic states of the molecule, while quantized collective oscillations of metallic conduction electrons are described by their classical plasma density. These approximations, as we have applied them within our minimal model, limit the possibility for charge transfer between molecule and metal, and effectively restrict our current presentation to the electromagnetic mechanism of SERS. We point out that a program similar to the minimal implementation of our approach has already been introduced in the time domain at the level of time-dependent HF theory for the molecular system and an electromagnetic boundary-element method for the particle \cite{Corni2001a}. Our approach is complementary and more general in the sense that we develop the full many-body theory starting from the exact many-body Hamiltonian for the interacting molecule-particle system and its interaction with an external electric field in the dipole approximation. Using a generalized M{\o}ller-Plesset perturbation theory \cite{Moller1934}, the effects of interaction with both particle and field are systematically and explicitly built into both nonperturbative and perturbative expressions for the molecular-electronic Green's function. Electronic-correlation effects beyond HF theory such as those of $n$th-order M{\o}ller-Plesset perturbation theory may be rigorously and straightforwardly included, while, alternatively, it is also clear how to treat the molecular-electronic sector of our theory within a Kohn-Sham formalism \cite{Kohn65}. Due to its general formulation, our approach recovers other approximate results from the literature, and, further, admits certain well-known observable features analytically upon invoking an analytic model for the particle's response. In particular, a closed-form expression for the quantum many-body SERS intensity, which is of the generic form
\begin{widetext}
\begin{equation*}
\frac{I_{\textrm{SERS}}({\bf k}')}{I_0({\bf k})}=\frac{\omega_{\bf k}\omega^3_{{\bf k}'}}{c^4}({\cal N}'+1)_{{\bf k}'\lambda'}\Big|\sum_{r;J}\hat{\bm\epsilon}^{(-)}_{\lambda'}({\bf k}')\cdot{\bm g}'_{rq}(-\hbar\omega_{{\bf k}'})\cdot\langle\nu_{J}^{\prime}|({\bf Q}_J-{\bf Q}_{0})\cdot\nabla_{{\bf Q}_J}\widetilde{\bm\alpha}^M_{pq,rr}(\hbar\omega_{\bf k},-\hbar\omega_{{\bf k}'})|\nu_{J}\rangle\cdot{\bm g}_{rq}(\hbar\omega_{{\bf k}})\cdot\hat{\bm\epsilon}^{(+)}_{\lambda}({\bf k})\Big|^2
\end{equation*}
\end{widetext}
with polarizability transition moments $\widetilde{\bm\alpha}_M,$ normal mode coordinates ${\bf Q}_J,$ and incident and Raman-scattered enhancement factors ${\bm g}$ and ${\bm g}',$ is developed and presented below in Eq. (\ref{IQSERS}). To our knowledge, this is the first place in the literature where a SERS intensity is derived, entirely from first principles, that explicitly treats the coupling and back reaction of a quantum molecular-electronic system with a nearby metallic particle in the presence of external perturbing radiation.

In Sec. \ref{CSERS}, we review an early and insightful classical model of SERS based upon the coupling and interaction of two dipoles with each other and with the external electric field. Two basic and essential types of interaction, the image effect and the local-field effect, are discussed and used to motivate our quantum-mechanical generalization. Starting from the exact molecule-particle Hamiltonian, the quantum many-body theory of SERS is developed in Sec. \ref{QSERS} where the effects of interaction of molecular electrons with metallic conduction electrons and with the external electric field are built into the underlying molecular-electronic Green's functions. The former effect, which accounts for the repeated interaction of the molecule with its own image, is included to infinite order in Sec. \ref{mep}, while the latter effect, which describes the interaction of the molecule with the local electric field of the particle, is included perturbatively in Sec. \ref{eep}. The random-phase approximation of the particle's polarizability is invoked in Sec. \ref{RPAsec} in order to demonstrate certain key properties analytically. Connection between the Green's function and scattering $T$-matrix is reviewed in Sec. \ref{TMATsec0}, and, subsequently, allows for the quantum-mechanical normal Raman-scattering intensity and enhanced Raman-scattering intensity to be computed in Secs. \ref{TMATsec1} and \ref{TMATsec2} respectively. Equation (\ref{IQSERS}), which displays a first-principles quantum-mechanical expression for the SERS intensity, is a major result of our work. Lastly, in Sec. \ref{LRT}, linear-response theory is reviewed and used to compute the induced density of the interacting molecular system in Sec. \ref{LRT2} and its influence upon the dynamics of the conduction electrons of the particle in Sec. \ref{PRP1}. Two appendices are devoted to the inclusion of molecular electron-electron interaction effects with density-functional theory, and to the Green's function based definition of the polarization propagator and linear polarizability.

Summation is implied over all repeated Greek indices. All integrals of the form $\int d^3x$ are taken over the volume of all space, while those of the form $\int d^4x\equiv\int d^3xdt$ are taken over the volume of all space and over all times from negative to positive infinity. Further, unless otherwise indicated, all time and frequency integrals run from negative to positive infinity. Molecular electronic state labels $i,j,k,l,\ldots$ refer to occupied or hole states, labels $a,b,c,d,\ldots$ refer to unoccupied or particle states, while the labels $p,q,r,s,\ldots$ are reserved for unspecified states.

\section{\label{CSERS}Review of Classical Model of SERS}
\begin{figure}[t]
%\psfrag{R}[][]{\rotatebox{12}{{\LARGE ${\bf x}_1-{\bf x}_2=r\hat{\bf r}$}}}
%\psfrag{p}[][]{{\LARGE ${\bf p}^{(1)}$}}
%\psfrag{m}[][]{{\LARGE ${\bf d}^{(1)}$}}
%\psfrag{E}[][]{{\LARGE ${\bf E}_0$}}
\begin{center}
%\rotatebox{0}{\resizebox{!}{4cm}{\includegraphics{dipoles}}}
\rotatebox{0}{\resizebox{!}{4cm}{\includegraphics{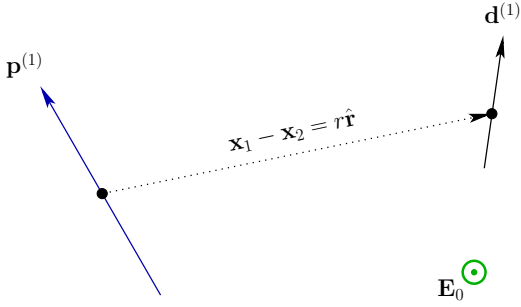}}}
\caption{\label{dipoles0} (Color online) In the SERS process, an incident external electric field ${\bf E}_0$ induces first-order molecular and particle dipole moments ${\bf d}^{(1)}$ and ${\bf p}^{(1)},$ which rebroadcast the incident radiation and, in doing so, couple to each other. Such a picture, which includes the reaction and back reaction of the dipoles to the incident field and among themselves, provides a insightful classical description of SERS.}
\end{center}
\end{figure}

Enhanced Raman scattering from a molecule can already be described to some extent at a classical level of theory. In 1980, Gersten and Nitzan \cite{nitzan1980} proposed a simple model consisting of two interacting electric dipoles: one dipole ${\bf d}^{(1)}$ representing a molecule and the other ${\bf p}^{(1)}$ representing an arbitrary polarizable body (taken here to be a metal particle) located nearby; see Fig. \ref{dipoles0}.
%!!!!!!!!!!!!!!!!!!!!!!!!!!!!!!!!!!!!!!!!!!!!!!!!!!!!!!!!!!!
%!!!!!!!!!!!!!!!!!!!!!!!!!!!!!!!!!!!!!!!!!!!!!!!!!!!!!!!!!!!
%!!!!!!!!!!!!!!!!!!!!!!!!!!!!!!!!!!!!!!!!!!!!!!!!!!!!!!!!!!!
%!!!!!!!!!!!!!!!!!!!!!!!!!!!!!!!!!!!!!!!!!!!!!!!!!!!!!!!!!!!
Both dipoles are induced, at first order, by an external electric field ${\bf E}_0$ that varies harmonically in time with frequency $\omega_{\bf k}$ and have corresponding molecular and particle polarizabilities $\widetilde{\bm\alpha}_M\equiv\widetilde{\bm\alpha}_M[{\bf Q}]$ and $\widetilde{\bm\alpha}_P,$ where $\bf Q$ labels a nuclear vibrational coordinate of the molecule; $\widetilde{\bm\alpha}_M$ and $\widetilde{\bm\alpha}_P$ may additionally depend upon frequency; tildes denote Fourier inversion to the frequency domain. Additionally, within the dipole approximation, both dipoles feel the effects of the incident radiation as rebroadcasted through electric dipole fields emanating from each other. In symbols, that is
\begin{equation}
\label{dips}
\begin{split}
{\bf d}^{(1)}&=\widetilde{\bm\alpha}_M\cdot({\bf E}_0+{\bf E}_P)\\
{\bf p}^{(1)}&=\widetilde{\bm\alpha}_P\cdot({\bf E}_0+{\bf E}_M),
\end{split}
\end{equation}
where the electric dipole fields $E_P^\xi=\Lambda^{\xi\sigma}p^{(1)}_\sigma$ and $E_M^\xi=\Lambda^{\xi\sigma}d^{(1)}_\sigma$ are expressed in terms of the matrix elements $\Lambda^{\xi\sigma}=[3\hat r^\xi\hat r^\sigma-\delta^{\xi\sigma}]/r^3$ of the second-rank dipole tensor ${\bm\Lambda}$ ($\xi,\sigma=x,y,z$). It is assumed here and throughout this article that the distance $r=|{\bf x}_1-{\bf x}_2|$ between molecule and particle is much less than the wavelength associated with ${\bf E}_0.$

In this way, enhanced Raman scattering from the coupled system of dipoles is achieved by allowing ${\bf d}^{(1)}$ and ${\bf p}^{(1)}$ in Eq. (\ref{dips}) to couple to and reach self-consistency with each other. From the total dipole moment ${\bf d}^{(1)}+{\bf p}^{(1)}=\widetilde{\bm\alpha}_{\textrm{tot}}\cdot{\bf E}_0,$ the total system polarizability
\begin{equation}
\begin{split}
\widetilde{\bm\alpha}_{\textrm{tot}}&=({\bf 1}-\widetilde{\bm\alpha}_M\cdot{\bm\Lambda}\cdot\widetilde{\bm\alpha}_P\cdot{\bm\Lambda})^{-1}\cdot\widetilde{\bm\alpha}_M\cdot({\bf 1}+{\bm\Lambda}\cdot\widetilde{\bm\alpha}_P)\\
&\ \ \ +({\bf 1}-\widetilde{\bm\alpha}_P\cdot{\bm\Lambda}\cdot\widetilde{\bm\alpha}_M\cdot{\bm\Lambda})^{-1}\cdot\widetilde{\bm\alpha}_P\cdot({\bf 1}+{\bm\Lambda}\cdot\widetilde{\bm\alpha}_M)
\end{split}
\end{equation}
may be directly obtained. The inverse matrix $({\bf 1}-\widetilde{\bm\alpha}_M\cdot{\bm\Lambda}\cdot\widetilde{\bm\alpha}_P\cdot{\bm\Lambda})^{-1}$ accounts for the classical {\it image effect} where the molecule's dipole moment repeatedly interacts through the particle with its own image dipole, while the term ${\bm\Lambda}\cdot\widetilde{\bm\alpha}_P$ (recall ${\bf E}_P\sim{\bm\Lambda}\cdot\widetilde{\bm\alpha}_P\cdot{\bf E}_0$) accounts for the {\it local-field effect} of the particle's dipole electric field upon the molecule. The second term accounts for the analogous image and local-field interactions where the roles of molecule and particle are reversed. Both of these effects will be generalized to a quantum-mechanical framework in the following.

The enhanced Raman polarizability 
\begin{equation}
\widetilde{\bm\alpha}_{\textrm{SERS}}=\Delta Q(\partial/\partial Q)\widetilde{\bm\alpha}_{\textrm{tot}}
\end{equation}
is related to $\widetilde{\bm\alpha}_{\textrm{tot}}$ by differentiation along the coordinate $\bf Q$ with amplitude $Q$ \cite{Placzek34,Albrecht1961}. After some algebra, it is found that the enhanced Raman scattering intensity in the direction $\hat{\bf k}'$ with polarization $\lambda'$ from the coupled system of dipoles is composed of the product of two terms: a normal Raman scattering intensity from the molecular dipole ${\bf d}^{(1)}$ in the absence of the particle dipole ${\bf p}^{(1)},$ and an enhancement factor stemming from the self-consistent coupling of the two dipoles, {\it i.e.},
\begin{equation}
\label{enh}
\begin{split}
\frac{I_{\textrm{SERS}}({\bf k}')}{I_0({\bf k})}&=\frac{\omega_{{\bf k}}\omega^3_{{\bf k}'}}{c^4}({\cal N}'+1)_{{\bf k}'\lambda'}\big|\hat{\bm\epsilon}_{\lambda'}({\bf k}')\cdot\widetilde{\bm\alpha}_{\textrm{SERS}}\cdot\hat{\bm\epsilon}_\lambda({\bf k})\big|^2\\
&=\frac{I_{\textrm{Raman}}({\bf k}')}{I_0({\bf k})}\big|\textrm{enhancement factor}\big|^2,
\end{split}
\end{equation}
where the normal Raman scattering intensity 
\begin{equation}
\label{I1}
\frac{I_{\textrm{Raman}}({\bf k}')}{I_0({\bf k})}=\frac{\omega_{{\bf k}}\omega^3_{{\bf k}'}}{c^4}({\cal N}'+1)_{{\bf k}'\lambda'}\Big|\hat{\epsilon}^\xi_{\lambda'}({\bf k}')\Delta Q\frac{\partial\widetilde{\alpha}_M^{\xi\sigma}}{\partial Q}\hat{\epsilon}^\sigma_\lambda({\bf k})\Big|^2.
\end{equation}
Here, ${\bf k}$ and ${\bf k}'$ are the wave vectors of the incident and Raman scattered electric fields with associated frequencies $\omega_{{\bf k}}$ and $\omega_{{\bf k}'}$ and polarizations $\hat{\bm\epsilon}_{\lambda}({\bf k})$ and $\hat{\bm\epsilon}_{\lambda'}({\bf k}'),$ where $\lambda,\lambda'=1,2.$ The intensity of the incident field is denoted by $I_0({\bf k})$ and ${\cal N}'_{{\bf k}'\lambda}$ is the number of Raman-scattered photons in the direction ${\bf k}'$ with polarization $\lambda'.$ For the purpose of computing the intensities in Eqs. (\ref{enh}) and (\ref{I1}), we have taken the amplitude of ${\bf E}_0$ equal to $i\sqrt{2\pi\hbar\omega_{\bf k}{\cal N}_{{\bf k}\lambda}/L^3}\hat{\bm\epsilon}_{\lambda}({\bf k}),$ where $L^3$ is the quantization volume of the electric field. This {\it ad hoc} choice was taken in anticipation of comparison to later results.

In the limit where the molecular and particle dipole moments are aligned with each other and with ${\bf E}_0$ and point along the $z$-axis, the enhancement factor in Eq. (\ref{enh}) is greatly simplified and takes the form \cite{Schatz-unp}
\begin{equation}
\label{cenh1}
\begin{split}
\big|\textrm{enhancement factor}\big|^2&=\Big|\frac{[1+(2/r^3)\widetilde\alpha_P]^2}{[1-\widetilde\alpha_P\widetilde\alpha_M(2/r^3)^2]^2}\Big|^2\\
&\approx\big|1+(2/r^3)\widetilde\alpha_P\big|^4,
\end{split}
\end{equation}
where it is assumed that $\omega_{{\bf k}}=\omega_{{\bf k}'},$ $\widetilde{\bm\alpha}_M$ and $\widetilde{\bm\alpha}_P$ are isotropic, and, in the second line, that $\widetilde\alpha_P\sim r^3$ and $\widetilde\alpha_M\ll r^3.$ This result, which is a straightforward extension of the ideas presented in Ref. \cite{nitzan1980}, demonstrates the characteristic fourth-power behavior of the SERS enhancement arising whenever a second polarizable body is able to react to and act back upon the first: with the largest enhancements occurring at a resonance of $\widetilde\alpha_P.$ Here, it should be pointed out that the SERS enhancement observed in experiment would scale in the same way if both the incident frequency and Raman-scattered frequency were the same. Further, it is clear that any quantum-mechanical generalization of the SERS mechanism should reduce in the appropriate limit to the classical results reviewed here.

\section{\label{QSERS}Quantum-Mechanical Approach to SERS}
The quantum many-body problem of SERS, consisting of a molecular-electronic system adsorbed to a metal particle having dimensions between tens and a few hundred nanometers and containing hundreds of thousands of atoms, poses a difficult theoretical and computational challenge. Response of such a system to an external perturbation presents one example where the associated properties are often sufficiently average to permit, to lowest order, a first-principles theoretical description \cite{thouless64}. It is precisely this situation that we now study in detail.

\subsection{Green's functions and many-body Hamiltonian of coupled molecule-particle system}
Basic to the theory of response of a zero-temperature quantum-mechanical many-body system to an external perturbing field is the one-body Green's function \cite{Martin59,Zubarev60,LO,FW}
\begin{equation}
i{\cal G}({\bf x},t;{\bf x}',t')=\langle\Phi^N_0|T\{\hat\Psi({\bf x},t)\hat\Psi^\dagger({\bf x}',t')\}|\Phi^N_0\rangle,
\end{equation}
defined as the expectation value of the time-ordered product of Heisenberg field operators $\hat\Psi$ and $\hat\Psi^\dagger$ in the normalized and interacting many-particle ground state $|\Phi^N_0\rangle.$ It measures the probability amplitude for an extra particle (or hole) to propagate under the action of the full many-body Hamiltonian $\hat H=\hat h+\hat V$ through an interacting assembly of $N$ particles from the space-time point $({\bf x}',t')$ to a later (earlier) point $({\bf x},t),$ there being found in the ground state $|\Phi^N_0\rangle.$ We note that application of Green's function methods to finite, nonuniform systems such as molecules \cite{Ohrn1965,Cederbaum1971,Ced1977}, atoms \cite{Reinhardt1969,Reinhardt1972a}, and atomic nuclei \cite{thouless64,Ring} have been extensively studied in the literature; see, {\it e.g.}, Ref. \cite{Ortiz1999} for a modern account.

Here we focus on the many-body SERS problem consisting of a Raman-active molecular system coupled to a nearby metallic particle supporting collective oscillation of its conduction electrons driven by an external perturbing field. As we are interested only in the dynamics of the molecular electrons in this article, we henceforth specialize to the case of fermion statistics where $\hat\Psi$ and $\hat\Psi^\dagger$ are electron field operators satisfying standard anticommutation relations at equal times, and $|\Phi^N_0\rangle$ is the exact $N$-electron ground state normalized to unity. Effects of electronic spin do not alter our proposed methodology in any significant way; we therefore omit the spin degree of freedom from all equations for simplicity.

In order to describe the quantum-mechanical interaction between molecular electrons and a metal particle illuminated by an external electric field, it will be necessary to introduce additional quantum fields for photons and for the quantized collective excitations of conduction electrons in the metal, known as plasmons. However, it is well known that the basic features of the electromagnetic enhancement mechanism of SERS can be explained by a classical description of the metal and field; see, {\it e.g.}, Ref. \cite{Schatz2007a}. In light of this fact, it is our desire to build a theory where both the conduction electrons in the metal particle and the external field are treated classically, yet the molecular-electronic system remains quantum mechanical. Henceforth, we assume that the quantized electric field $\hat{\bf E}_0$ is well approximated by the classical field ${\bf E}_0(t)={\bf E}^{(+)}_{0}(t)+{\bf E}^{(-)}_{0}(t),$ and we further assume a Bogoliubov decomposition \cite{Bogolubov1947a} of the quantized plasmon field
\begin{equation}
\label{bogo}
\hat\Omega({\bf x},t)=\sqrt{n({\bf x},t)}+\delta\hat\Omega({\bf x},t)
\end{equation}
in terms of the classical conduction electron density $n$ and small-amplitude quantum fluctuations $\delta\hat\Omega$ around $n.$ The specific form of this classical component will be expounded upon in the following. We note that $\int n({\bf x},t)d^3x$ is approximately equal to the total number of conduction electrons participating in collective excitation; nonetheless, such excitations will continue to be called plasmons in spite of their representation by a classical field.

In terms of $\hat\Psi$ and $\hat\Omega,$ the many-body Hamiltonian of the interacting molecular electron-plasmon assembly may be written in second quantization as
\begin{widetext}
\begin{equation}
\label{Hsimp}
\begin{split}
\hat{H}_{\textrm{tot}}&=V_{N}+{\tint}\hat{\Psi}^\dagger({\bf x},t)h({\bf x})\hat{\Psi}({\bf x},t)d^{3}x+{\tint}\hat{\Omega}^\dagger({\bf x},t)h({\bf x})\hat{\Omega}({\bf x},t)d^{3}x+({1}/{2}){\tint}\hat{\Psi}^\dagger({\bf x},t)\hat{\Psi}^\dagger({\bf x}',t)V({\bf x},{\bf x}')\hat{\Psi}({\bf x}',t)\hat{\Psi}({\bf x},t)d^{3}xd^{3}x'\\
&\ \ \ +({1}/{2}){\tint}\hat{\Omega}^\dagger({\bf x},t)\hat{\Omega}^\dagger({\bf x}',t)V({\bf x},{\bf x}')\hat{\Omega}({\bf x}',t)\hat{\Omega}({\bf x},t)d^{3}xd^{3}x'+{\tint}\hat{\Psi}^\dagger({\bf x},t)\hat{\Omega}^\dagger({\bf x}',t)V({\bf x},{\bf x}')\hat{\Omega}({\bf x}',t)\hat{\Psi}({\bf x},t)d^{3}xd^{3}x',
\end{split}
\end{equation}
\end{widetext}
where a Born-Oppenheimer separation has been made between electron and nuclear coordinates \cite{BOapprox} and all off-diagonal terms in the field operators are omitted. Here, $V_N\equiv V_N(\{{\bf R}\})$ is the nuclear-nuclear repulsion energy which depends upon the set $\{{\bf R}\}\equiv\{{\bf R}_1,\ldots,{\bf R}_M\}$ of all $M$ nuclear coordinates; note that the nuclear kinetic energy $T_N\equiv T_N(\{{\bf R}\})$ does not appear in Eq. (\ref{Hsimp}) as $\hat H$ describes only the molecular electron-plasmon physics. Terms two and three, involving $h,$ represent the single-particle Hamiltonians for noninteracting electron and plasmon systems, including the external electron-nuclear attraction, perturbed by a common externally applied classical electric field ${\bf E}_0$ within the dipole interaction approximation. Introducing molecular and particle dipole moments $\hat {\bf d}(t)=\int\hat{\Psi}^\dagger({\bf x},t)(-e{\bf x})\hat{\Psi}({\bf x},t)d^3x$ and ${\bf p}(t)\approx\int(-e{\bf x})n({\bf x},t)d^3x$ respectively, this perturbation takes the standard form $-\hat {\bf d}\cdot{\bf E}_0$ and $-{\bf p}\cdot{\bf E}_0,$ where the electric field is evaluated at the molecular-frame origin (${\bf x}_1={\bf 0}$). Terms four and five represent the potential energy of pairwise repulsive interaction between electrons and, separately, plasmons; the last term expresses the potential energy of interaction between molecular electrons and plasmons. The two-body potential $V({\bf x},{\bf x}')=e^2/|{\bf x}-{\bf x}'|$ appearing above is the instantaneous Coulomb interaction with electronic charge $-e.$

We now invoke a second Born-Oppenheimer-like separation between molecular electrons and metallic conduction electrons; here we consider only the molecular part of the Hamiltonian in Eq. (\ref{Hsimp}) including its interaction with the external radiation field and metallic conduction electrons. Effects stemming from the noninteracting plasmon Hamiltonian (term three) as well as the plasmon-plasmon interaction (term five) are implicitly accounted for through the dynamics of the particle's induced electronic density, which evolves under the action of 
\begin{equation}
\label{partH}
\begin{split}
&\hat H_P={W}+{\tint}\hat{\Omega}^\dagger({\bf x},t)h({\bf x})\hat{\Omega}({\bf x},t)d^{3}x\\
&+({1}/{2}){\tint}\hat{\Omega}^\dagger({\bf x},t)\hat{\Omega}^\dagger({\bf x}',t)V({\bf x},{\bf x}')\hat{\Omega}({\bf x}',t)\hat{\Omega}({\bf x},t)d^{3}xd^{3}x',
%&+{\tint}\hat{\Omega}^\dagger({\bf x},t)\hat{\Psi}^\dagger({\bf x}',t)V({\bf x},{\bf x}')\hat{\Psi}({\bf x}',t)\hat{\Omega}({\bf x},t)d^{3}xd^{3}x'.
\end{split}
\end{equation}
where ${W},$ defined below in Eq. (\ref{MPES}), is the interacting molecular ground-state potential energy dependent upon the underlying classical plasmon density $n.$ The associated response of the particle is detailed below in Sec. \ref{PRP1}. Eliminating these terms leaves the simplified molecular Hamiltonian
\begin{widetext}
\begin{equation}
\label{simpHAM}
\begin{split}
\hat{H}&=V_N+{\tint}\hat{\Psi}^\dagger({\bf x},t)h({\bf x})\hat{\Psi}({\bf x},t)d^{3}x+{\tint}V({\bf x},{\bf x}')\hat{\Psi}^\dagger({\bf x},t)[(1/2)\hat{\Psi}^\dagger({\bf x}',t)\hat{\Psi}({\bf x}',t)+\hat{\Omega}^\dagger({\bf x}',t)\hat{\Omega}({\bf x}',t)]\hat{\Psi}({\bf x},t)d^3xd^{3}x'\\
&\approx V_N+{\tint}\hat{\Psi}^\dagger({\bf x},t)h({\bf x})\hat{\Psi}({\bf x},t)d^{3}x+{\tint}V({\bf x},{\bf x}')\hat{\Psi}^\dagger({\bf x},t)[(1/2)\hat{\Psi}^\dagger({\bf x}',t)\hat{\Psi}({\bf x}',t)+n({\bf x}',t)]\hat{\Psi}({\bf x},t)d^3xd^{3}x',
%&={\tint}d^{3}x\hat{\Psi}^\dagger({\bf x},t)\Bigl\{h({\bf x})+{\tint}d^{3}x'V({\bf x},{\bf x}')[(1/2)\hat{\Psi}^\dagger({\bf x}',t)\hat{\Psi}({\bf x}',t)+\hat{\Omega}^\dagger({\bf x}',t)\hat{\Omega}({\bf x}',t)\Bigr]\Bigr\}\hat{\Psi}({\bf x},t).
%&={\tint}d^{3}x\hat{\Psi}^\dagger({\bf x},t)h({\bf x})\hat{\Psi}({\bf x},t)+({1}/{2}){\tint}d^{3}xd^{3}x'\hat{\Psi}^\dagger({\bf x},t)\hat{\Psi}^\dagger({\bf x}',t)V({\bf x},{\bf x}')\hat{\Psi}({\bf x}',t)\hat{\Psi}({\bf x},t)+{\tint}d^{3}xd^{3}x'\hat{\Psi}^\dagger({\bf x},t)\hat{\Omega}^\dagger({\bf x}',t)V({\bf x},{\bf x}')\hat{\Omega}({\bf x}',t)\hat{\Psi}({\bf x},t).
\end{split}
\end{equation}
\end{widetext}
where the plasmon field operator is expanded to lowest order and the products $\hat{\Omega}^\dagger\hat{\Omega}$ are replaced by the classical conduction electron density $n.$ Here, $h=h_0-(-e{\bf x})\cdot{\bf E}_0$ and $h_0$ contains the molecular-electronic kinetic energy and external electron-nuclear attraction.

The Heisenberg electron field operator $\hat{\Psi}({\bf x},t)=\exp(i\hat Ht/\hbar)\hat{\Psi}({\bf x})\exp(-i\hat Ht/\hbar)$ with $\hat{\Psi}({\bf x})=\sum_p\phi_p({\bf x})\hat c_p$ may now be expanded onto the basis of electron annihilation operators 
\begin{equation}
%\begin{split}
\hat c_p=(1-\rho_p)\hat a_p+\rho_p\hat b^\dagger_p
%\hat c_p^\dagger(t)&=[1-\rho_p(t)]\hat a_p^\dagger(t)+\rho_p(t)\hat b_p(t).
%\end{split}
\end{equation}
expressed in terms of the particle and hole operators $\hat a_p$ and $\hat b_p$ through canonical transformation. These fermionic operators all satisfy standard anticommutation relations. The occupation numbers $\rho_p,$ which are eigenvalues of the exact one-body reduced density
\begin{equation}
%\begin{split}
\rho({\bf x},{\bf x}')=\langle\Phi^N_0|\Psi^\dagger({\bf x})\Psi({\bf x}')|\Phi^N_0\rangle=\sum_{pq}\phi_p^*({\bf x})\rho_{pq}\phi_q({\bf x}')
%\end{split}
\end{equation}
with matrix elements $\rho_{pq}=\langle\Phi^N_0|\hat c_p^\dagger\hat c_q|\Phi^N_0\rangle,$ take values of 0 or 1 depending on the occupation of the $p$th single-particle state.

In terms of these basic operators and one-body wave functions (molecular orbitals) $\phi_p({\bf x})=\langle {\bf x}|\phi_p\rangle\equiv\langle {\bf x}|p\rangle=\langle {\bf x}|\hat c_p^\dagger|{\textrm{vac}}\rangle$ with physical vacuum $|{\textrm{vac}}\rangle,$ the electronic Hamiltonian reduces to
\begin{equation}
\label{simpH}
\hat H=V_N+\sum_{pq}\langle p|h+U|q\rangle\hat c^\dagger_p\hat c_q+\frac{1}{2}\sum_{pqrs}\langle pq|V|rs\rangle\hat c^\dagger_p\hat c^\dagger_q\hat c_s\hat c_r,
\end{equation}
where the sum runs over all $N$ electrons of the molecular system, and the explicit time dependence of the external fields have been omitted for simplicity of notation. The molecular dipole moment is expanded as $\hat{\bf d}=\sum_{pq}\langle p|-e{\bf x}|q\rangle\hat c_p^\dagger\hat c_q$ and included within the one-electron Hamiltonian $h.$ Here, $U({\bf x},t)=\int V({\bf x},{\bf x}')n({\bf x}',t)d^3x'$ is the plasmon potential with matrix elements
\begin{equation}
\label{U}
\langle p|U(t)|q\rangle={\tint}d^3x'\langle p|V|q\rangle({\bf x}')n({\bf x}',t).
\end{equation}
The collective electronic density of the particle
\begin{equation}
\label{density}
%\begin{split}
%n_{\textrm{ext}}({\bf x},t)&\approx n_0({\bf x})+n^{(1)}({\bf x},t)\\
n({\bf x},t)=n_0({\bf x})+\delta n({\bf x},t)
%\end{split}
\end{equation}
may be decomposed into the sum of a static density $n_0$ and low-amplitude excitations $\delta n\equiv n_{\textrm{int}}$ induced by some interaction to be specified later. Similarly, $U({\bf x},t)=U_0({\bf x})+U_{\textrm{int}}({\bf x},t)$ factors into a static potential $U_0$ and an induced potential $U_{\textrm{int}}$ associated with $n_0$ and $n_{\textrm{int}}$ respectively.

Application of Wick's theorem reduces the above electronic Hamiltonian (\ref{simpH}) to the sum of three parts: a scalar constant
\begin{equation}
\label{E0}
E_0=V_N+\sum_k\langle k|h_0+e{\bf x}\cdot{\bf E}_0+U|k\rangle+\frac{1}{2}\sum_{kl}\langle kl|V|kl-lk\rangle,
\end{equation}
a one-body term quadratic in electron operators
\begin{equation}
\begin{split}
\label{Fock_op}
\hat F&=\sum_{pqk}\bigl[\langle p|h_0+e{\bf x}\cdot{\bf E}_0+U|q\rangle+\langle pk|V|qk-kq\rangle\bigr] N\{\hat c_p^\dagger\hat c_q\}\\
&=\hat F_0+\sum_{pq}\langle p|e{\bf x}\cdot{\bf E}_0+U_{\textrm{int}}|q\rangle N\{\hat c_p^\dagger\hat c_q\},
\end{split}
\end{equation}
and a two-body term involving the product of four electron operators
\begin{equation}
\label{hint}
\hat H_{\textrm{int}}=\frac{1}{2}\sum_{pqrs}\langle pq|V|rs\rangle N\{\hat c^\dagger_p\hat c^\dagger_q\hat c_s\hat c_r\}.
\end{equation}
In the above, normal ordering is taken with respect to particles and holes ({\it{i.e.}}, $\hat a_p$ and $\hat b_q$) rather than basic electron operators $\hat c_p.$

The quadratic term $\hat F$ in Eq. (\ref{Fock_op}) is decomposable into the sum of the Fock operator $\hat F_0=\sum_{pq}\langle p|F_0|q\rangle N\{\hat c_p^\dagger\hat c_q\}$ and a one-body electron-field and electron-plasmon interaction
\begin{equation}
\label{Fint}
\hat F_{\textrm{int}}=\sum_{pq}\langle p|e{\bf x}\cdot{\bf E}_0+U_{\textrm{int}}|q\rangle N\{\hat c_p^\dagger\hat c_q\}.
\end{equation}
Choosing the underlying molecular orbitals $\phi_p$ to diagonalize the Fock matrix $\langle p|F_0|q\rangle\equiv\langle p|h_0+U_0|q\rangle+\sum_k\langle pk|V|qk-kq\rangle=\delta_{pq}\varepsilon^0_p,$ yields the familiar HF orbital equation with effective potential $U_0$
\begin{equation}
\label{HFE}
\begin{split}
[h_0(&{\bf x})+U_0({\bf x})]\phi_q({\bf x})\\
&+\sum_k[\langle k|V|k\rangle({\bf x})\phi_q({\bf x})-\langle k|V|q\rangle({\bf x})\phi_k({\bf x})]=\varepsilon^0_q\phi_q({\bf x})
%\varepsilon^0_q&\phi_q({\bf x})=\\
\end{split}
\end{equation}
for $\phi_q$ and with orbital energy $\varepsilon^0_q=\langle q|h_0+U_0|q\rangle+\sum_k\langle qk|V|qk-kq\rangle.$ Here, and throughout, $\langle pq|V|rs-sr\rangle\equiv\langle pq|V|rs\rangle-\langle pq|V|sr\rangle$ are antisymmetrized matrix elements of the two-body potential $V.$

Through this normal ordering, the electronic Hamiltonian in Eq. (\ref{simpHAM}) now takes the M{\o}ller-Plesset form \cite{Ostlund}
\begin{equation}
\hat H=E_0+\hat F_0+\hat F_{\textrm{int}}+\hat H_{\textrm{int}},
\end{equation}
where $\hat F_0=\sum_p\varepsilon^0_pN\{\hat c_p^\dagger\hat c_p\}=\sum_a\varepsilon^0_a\hat a_a^\dagger\hat a_a-\sum_k\varepsilon^0_k\hat b_k^\dagger\hat b_k$ is diagonal when expressed in the HF basis. Associated with it are the formally exact ground-state energy ${W}$ and ground state vector $|\Phi^N_0\rangle$ stemming from the interacting molecular Schr\"odinger equation
\begin{equation}
\label{MPES}
\hat H|\Phi^N_0\rangle=[E_0+\hat F_0+\hat F_{\textrm{int}}+\hat H_{\textrm{int}}]|\Phi^N_0\rangle={W}|\Phi^N_0\rangle.
\end{equation}
The ground-state energy ${W}$ depends parametrically upon the underlying classical plasmon density $n.$ Separation of $\hat H$ into an unperturbed component $\hat H_0=E_0+\hat F_0$ and a perturbation $\hat F_{\textrm{int}}+\hat H_{\textrm{int}}$ provides a natural ansatz for the application of many-body perturbation theory (see, {\it e.g.}, Ref. \cite{Ostlund}). This will be the subject of Sec. \ref{MBPT1}. 

%We note that the above reorganization has effectively renormalized $V_N\to E_0$ and $\langle p|h_0|q\rangle\to\langle p|F_0|q\rangle$ in the many-body Hamiltonian (\ref{simpH}).

\subsection{\label{ip}Interaction picture}
We now adopt the interaction picture with respect to the uncorrelated noninteracting $N$-electron HF ground state $|\Phi^N_{\textrm{HF}}\rangle=\prod_p\hat c^\dagger_p|\textrm{vac}\rangle$ at zero temperature; it satisfies $\hat F_0|\Phi^N_{\textrm{HF}}\rangle=-\sum_k\varepsilon^0_k|\Phi^N_{\textrm{HF}}\rangle$ and is normalized to unity. The reference state $|\Phi^N_{\textrm{HF}}\rangle$ is the Fermi vacuum for the electron creation and annihilation operators (and particle-hole operators). Henceforth, all expectation values will be computed within $|\Phi^N_{\textrm{HF}}\rangle.$ The molecular orbitals $\phi_p$ underlying $|\Phi^N_{\textrm{HF}}\rangle$ are determined self-consistently and satisfy $\langle p|q\rangle=\delta_{pq}$; it will assumed that both $|\Phi^N_{\textrm{HF}}\rangle$ and the corresponding orbitals and orbital energies are known.

%, {\it i.e.}, $\mu=\varepsilon^0_F.$
%It is referred to as noninteracting because each of its $N$ electrons occupies its own independent single-particle state and moves in the mean field generated by the remaining ${\cal N}-1$ electrons plus the external field of the atomic nuclei and effective potential $U_0$ of the metal particle as is evident from Eq. (\ref{HFE}); no real electron-electron, electron-field, nor electron-plasmon interactions are suffered. 

The zero-temperature HF one-body Green's function \cite{LO,FW} is defined with respect to the Fermi vacuum as
\begin{equation}
\begin{split}
\label{Gtime}
iG^{(0)}({\bf x},t;{\bf x}',t')&=\langle\Phi^N_{\textrm{HF}}|T\{\hat\Psi({\bf x},t)\hat\Psi^\dagger({\bf x}',t')\}|\Phi^N_{\textrm{HF}}\rangle\\
%&=\sum_{pq}\phi_p({\bf x})\langle\Phi^N_{\textrm{HF}}|T\{\hat c_p(t)\hat c_q^\dagger(t')\}|\Phi^N_{\textrm{HF}}\rangle\phi_q^*({\bf x}')\\
&=\sum_{pq}\phi_p({\bf x})iG_{pq}^{(0)}(t,t')\phi_q^*({\bf x}'),
\end{split}
\end{equation}
where $iG_{pq}^{(0)}(t,t')=\langle\Phi^N_{\textrm{HF}}|T\{\hat c_p(t)\hat c_q^\dagger(t')\}|\Phi^N_{\textrm{HF}}\rangle$ represents the Fock-space matrix elements of $iG^{(0)}$ and $\hat c_p(t)=\exp(i\hat F_0t/\hbar)\hat c_p\exp(-i\hat F_0t/\hbar)=\exp(-i\varepsilon^0_pt/\hbar)\hat c_p$ in the interaction-picture representation. The orbitals $\phi_q$ are solutions of the HF equations (\ref{HFE}). In Fourier space, it has the Lehmann spectral representation
\begin{equation}
\label{LG}
\begin{split}
&\widetilde G^{(0)}({\bf x},{\bf x}';\omega)=\sum_{pq}\phi_p({\bf x})\widetilde G^{(0)}_{pq}(\omega)\phi_q^*({\bf x}')\\
&=\sum_{pq}\phi_p({\bf x})\delta_{pq}\Bigl[\frac{1-\rho_p^0}{\omega+i0^+-\varepsilon_p^0/\hbar}+\frac{\rho_p^0}{\omega-i0^+-\varepsilon_p^0/\hbar}\Bigr]\phi_q^*({\bf x}'),
\end{split}
\end{equation}
where the eigenvalue $\rho_p^0$ of the HF one-matrix $\rho_{pq}=\langle\Phi^N_{\textrm{HF}}|\hat c_p^\dagger\hat c_q|\Phi^N_{\textrm{HF}}\rangle=\rho_p^0\delta_{pq}$ is the occupation number of the $p$th single-particle state and $0^+$ is a positive infinitesimal needed only to damp and subsequently converge certain Fourier integrals; $0^+$ should be taken to zero at the end of all calculations. The retarded component of the time-ordered Green's function $\widetilde G^{(0)}$
\begin{equation}
\label{retG}
[\widetilde G_{(0)}^{R}]_{pq}(\omega)=\frac{\delta_{pq}}{\omega+i0^+-\varepsilon_p^0/\hbar}
\end{equation}
will be of use in the following. In Eq. (\ref{Gtime}), $G_{pq}^{(0)}$ is the zeroth order propagator for an extra electron to propagate from the state $q$ to the state $p$ within the $N$-particle noninteracting background $|\Phi^N_{\textrm{HF}}\rangle.$ In this approximation, the electron does not interact directly with any other particle nor is it scattered out of the single-particle state $p$ (due to the delta function $\delta_{pq}$ in Eq. (\ref{LG})). Interaction effects will now be systematically built in through the machinery of time-dependent many-body perturbation theory \cite{Merz,FW}.

\subsection{\label{MBPT1}Many-body perturbation theory}
Until this point we have developed an essentially exact many-body theory of SERS including the coupling of a general molecule to a plasmon-supporting metallic system under the influence of an external perturbing electric field. Building from the noninteracting HF reference state of the molecule, it will now be demonstrated how to systematically incorporate these interaction effects to arbitrary order through the machinery of many-body perturbation theory. As a first step in this direction, we choose to consider only those interactions originating from the external electric field and from collective excitations of conduction electrons in the metal particle induced by both the external field and by interactions with the molecule, and truncate the level of molecular-electronic correlation to HF mean-field theory; therefore, we suppress the two-body electron-electron interaction $\hat H_{\textrm{int}}$ in Eq. (\ref{hint}) and focus upon the perturbations stemming from $\hat F_{\textrm{int}}.$ It is, however, important to note that, due to the generality of our approach, electronic-correlation effects (from $\hat H_{\textrm{int}}$) may be included by using the same perturbation techniques described below, or, alternatively, may be added in the spirit of a Kohn-Sham density-functional theory approach \cite{Kohn65}. Within this context we note that the chemical mechanism of SERS may be explored by including a subset of the metallic conduction electrons in addition to the molecular electrons within the density-functional theory; see Appendix \ref{appDFT} for details. 

%It will be assumed in the following that ${\bf p}_0={\bf 0}.$ We do not anticipate any significant changes in the proposed methodology, even when the particle has a permanent dipole moment.

\subsubsection{\label{mep}Interaction of molecular electrons with a metal particle}
We focus first on the perturbations of the molecular-electronic system induced by the presence of a metal particle nearby. From Eq. (\ref{density}), the collective electronic density of the particle 
\begin{equation}
n({\bf x},t)=n_0({\bf x})+n_M({\bf x},t)
\end{equation}
may be decomposed as the sum of a static density $n_0$ and low-amplitude excitations $\delta n\equiv n_M$ induced by the molecule itself. This will be shown to underlie the quantum-mechanical analog of the {\it image effect} discussed in Ref. \cite{nitzan1980}. Associated with these densities are the static and induced dipole moments ${\bf p}_0$ and ${\bf p}_M$ of the particle. With this decomposition, the interaction between molecular electrons and the induced density $n_M$ in the particle can be written to lowest order as
\begin{equation}
\begin{split}
F^{\textrm{int}}_{pq}(t)\to F^P_{pq}(t)&=\langle p|{\tint}d^3x'V({\bf x},{\bf x}')n_M({\bf x}',t)|q\rangle\\
&\approx-\langle p|-e{\bf x}|q\rangle\cdot\frac{3\hat{\bf r}[{\bf p}_M(t)\cdot\hat{\bf r}]-{\bf p}_M(t)}{r^3}
\end{split}
\end{equation}
following multipole expansion of the potential $V,$ where ${\bf x}_1-{\bf x}_2=r\hat{\bf r}$ and where the ${\bf p}_M(t)=\int(-e{\bf x})n_M({\bf x},t)d^3x$ is the dipole moment induced in the particle by the molecule. Effects of retardation are neglected. In principle there is no reason to additionally impose a dipole interaction approximation between molecule and particle; however, since we are interested here in developing a quantum-mechanical generalization of the classical dipole model presented in Sec. \ref{CSERS}, we choose to do so. Note that the interaction between molecular electrons and the static part of the density $n_0$ is already accounted for in the reference system; see, {\it e.g.}, Eq. (\ref{HFE}). The above expression represents the interaction energy between the molecular dipole moment $\hat{\bf d}$ and a molecule-induced excitation of conduction electrons in the metal particle. It can be rewritten in more compact form as
\begin{equation}
\label{UM}
\begin{split}
\hat F_{\textrm{int}}(t)\to\hat F_P(t)&=\sum_{pq}\langle p|U_P(t)|q\rangle N\{\hat c_p^\dagger(t)\hat c_q(t)\}\\
&\approx-\hat{\bf d}(t)\cdot{\bm\Lambda}\cdot{\bf p}_M(t),
%=\hat{\bf d}(t)\cdot{\bm\Lambda}\cdot{\tint}d^3x(-e{\bf x})n_M({\bf x},t),
%-\hat{\bf d}(t)\cdot{\bf E}_P(t)
\end{split}
\end{equation}
where, as before, $\Lambda^{\xi\sigma}=[3\hat r^\xi\hat r^\sigma-\delta^{\xi\sigma}]/r^3$ are the matrix elements of the second-rank dipole tensor ${\bm\Lambda}.$

Dyson has provided an algorithm for computing the exact interacting $N$-body Green's function by perturbative expansion from the noninteracting reference state of the interaction picture \cite{FW}, where the effects of interaction are encapsulated within the irreducible self energy $\hbar\Sigma^\bigstar$ \cite{star}. In the particular case of the perturbation $\hat F_P=-\hat{\bf d}\cdot{\bm\Lambda}\cdot{\bf p}_M,$ Dyson's expansion yields
\begin{widetext}
\begin{equation}
\label{link}
\begin{split}
&i{\cal G}_{pq}(t,t')=\sum_{n=0}^\infty\frac{(-i/\hbar)^n}{n!}{\tint}dt_1\cdots dt_n\langle\Phi^N_{\textrm{HF}}|T\{[-\hat{\bf d}(t_1)\cdot{\bm\Lambda}\cdot{\bf p}_M(t_1)\cdots-\hat{\bf d}(t_n)\cdot{\bm\Lambda}\cdot{\bf p}_M(t_n)]\hat c_p(t)\hat c_q^\dagger(t')\}|\Phi^N_{\textrm{HF}}\rangle_C\\
&=iG^{(0)}_{pq}(t,t')+\frac{i}{\hbar}\sum_{rs}{\tint}d^4x_1G^{(0)}_{pr}(t,t_1)\langle r|e{\bf x}|s\rangle\cdot{\bm\Lambda}\cdot(-e{\bf x}_1)n_M({\bf x}_1,t_1)G^{(0)}_{sq}(t_1,t')\\
&\ \ \ +\frac{i}{\hbar^2}\sum_{rstu}{\tint}d^4x_1d^4x_1'G^{(0)}_{pr}(t,t_1')\langle r|ex^\xi|t\rangle\Lambda^{\xi\sigma}(ex_1^\sigma)i\Pi_P({\bf x}_1,t_1;{\bf x}_1',t_1')(ex_1^{\prime\beta})\Lambda^{\beta\gamma}G^{(0)}_{tu}(t_1',t_1)\langle u|ex^{\prime\gamma}|s\rangle G^{(0)}_{sq}(t_1,t')+\cdots,
\end{split}
\end{equation} 
\end{widetext}
where $\langle\cdots\rangle_C$ indicates that only linked or connected diagrams contribute in the expansion \cite{Goldstone57}.

The classical polarization propagator of the molecule-induced density fluctuations $n_M=n-n_0$ in the particle can be identified in the second-order term above as
\begin{equation}
\label{piM}
i\Pi_P({\bf x},t;{\bf x}',t')=n_M({\bf x},t)n_M({\bf x}',t')
%=T\{n_M({\bf x},t)n_M({\bf x}',t')\}
\end{equation}
in analogy to the quantum-mechanical noninteracting HF polarization propagator \cite{LO,FW} defined in Eq. (\ref{HFPi}) in Appendix \ref{PP}. From the retarded component of $\Pi_P,$ the molecule-induced classical linear polarizability of the particle can be written as
\begin{equation}
\label{alphaP}
\begin{split}
-i\hbar\alpha_P^{\xi\sigma}(t,t')&={\tint}d^3xd^3x'(-ex^\xi)i\Pi^R_P({\bf x},t;{\bf x}',t')(-ex^{\prime\sigma})\\
%&={\tint}d^3xd^3x'(-ex^\xi)\theta(t-t')n_M({\bf x},t)n_M({\bf x}',t')(-ex^{\prime\sigma})\\
&=\theta(t-t')p_M^\xi(t)p_M^\sigma(t')
\end{split}
\end{equation} 
and has a mathematical structure similar to the noninteracting HF molecular linear polarizability (\ref{HFalph}) defined in Appendix \ref{PP}. Following omission of the first-order perturbative correction, which is of no importance in the following (and can be renormalized away), the retarded component of Eq. (\ref{alphaP}) becomes
\begin{widetext}
\begin{equation}
\begin{split}
{\cal G}^R_{pq}&(t,t')-[G^R_{(0)}]_{pq}(t,t')\\
&=\sum_{rs}{\tint}dt_1dt_1'[G^R_{(0)}]_{pr}(t,t_1)\Big[\frac{-1}{\hbar^2}\sum_{tu}\langle r|e{\bf x}|t\rangle\cdot{\bm\Lambda}\cdot i\hbar{\bm\alpha}_P(t_1,t_1')\cdot{\bm\Lambda}\cdot[G^R_{(0)}]_{tu}(t_1',t_1)\langle u|e{\bf x}'|s\rangle\Big][G^R_{(0)}]_{sq}(t_1,t')+\cdots
\end{split}
\end{equation} 
\end{widetext}
with irreducible self energy
\begin{equation}
\begin{split}
\hbar[\Sigma^\bigstar_R]_{pq}(t_1&,t_1')=\frac{-1}{\hbar}\sum_{rs}\langle p|e{\bf x}|r\rangle\cdot{\bm\Lambda}\\
&\cdot i\hbar{\bm\alpha}_P(t_1,t_1')\cdot{\bm\Lambda}\cdot[G^R_{(0)}]_{rs}(t_1',t_1)\langle s|e{\bf x}'|q\rangle+\cdots.
\end{split}
\end{equation} 
Here, the first term on the right hand side already contains the desired effects of polarization. We, therefore, truncate the above perturbation series for $\hbar\Sigma^\bigstar$ at second order in $\hat F_P$ and define the spectral representation of the second-order irreducible self energy by
\begin{equation}
\label{ISE}
\begin{split}
\hbar[\widetilde\Sigma_R^{\bigstar}]&^{(P,2)}_{pq}(\omega)=\sum_{rs}\langle p|-e{\bf x}|r\rangle\cdot{\bm\Lambda}\\
&\hspace{0.5cm}\cdot{\tint}\frac{d\omega'}{2\pi i}\widetilde{\bm\alpha}_P(\omega')[\widetilde G^R_{(0)}]_{rs}(\omega+\omega')\cdot{\bm\Lambda}\cdot\langle s|-e{\bf x}'|q\rangle.
\end{split}
\end{equation}
An approximate yet infinite-order nonperturbative representation of the exact interacting one-body retarded Green's function may now be constructed by resumming the Dyson series 
\begin{equation}
\label{DEGF}
\begin{split}
&[\widetilde{\cal G}^R_P]_{pq}(\omega)\\
&=[\widetilde G^R_{(0)}]_{pq}(\omega)+\sum_{rs}[\widetilde G^R_{(0)}]_{pr}(\omega)[\widetilde\Sigma^{\bigstar}_R]^{(P,2)}_{rs}(\omega)[\widetilde{\cal G}^R_P]_{sq}(\omega)
%[\widetilde{G}^R_{(0)}]_{sq}(\omega)+\cdots
%&=G^{(0)}_{pq}(t,t')+\sum_{q'r'}{\tint}dt_1G^{(0)}_{pr'}(t,t_1)\Big[\frac{1}{\hbar^2}\sum_{p's'}{\tint}dt_1'\langle r'|e{\bf x}|s'\rangle\cdot{\bm\Lambda}i{\bm\alpha}_M(t_1,t_1'){\bm\Lambda}\cdot G^{(0)}_{s'p'}(t_1',t_1)\langle p'|e{\bf x}'|q'\rangle\Big]G^{(0)}_{q'q}(t_1,t')
\end{split}
\end{equation} 
stemming from the second-order perturbative approximation $\widetilde\Sigma_R^{\bigstar}\approx[\widetilde\Sigma_R^{\bigstar}]^{(P,2)}$ \cite{CT1992}. Such an approximation corresponds to the physical scenario where the molecule is able to repeatedly excite density fluctuations in and polarize the particle and these excitations act back upon the molecule to infinite order. Said differently, the quantum analog of the classical image effect discussed in Ref. \cite{nitzan1980} is included by resumming the Dyson series based upon the above second-order truncation of the self energy. For simplicity of notation we drop the label 2 in the following so that $[\widetilde\Sigma^{\bigstar}_R]^P\equiv[\widetilde\Sigma^{\bigstar}_R]^{(P,2)}.$ Note that $[\widetilde\Sigma_R^{\bigstar}]^P$ contains only one of several irreducible terms occurring at second order in perturbation theory \cite{FW}; however, among all others, this is the only contribution that includes the desired polarization effects.

In similar spirit, the approximate time-ordered one-body Green's function $\widetilde{\cal G}^P$ may be compactly expressed by the Dyson expansion
\begin{equation}
\label{intG2}
\widetilde{\cal G}^P_{pq}(\omega)=\widetilde G^{(0)}_{pq}(\omega)+\sum_{rs}\widetilde G^{(0)}_{pr}(\omega)[\widetilde\Sigma^{\bigstar}_P]_{rs}(\omega)\widetilde{\cal G}^P_{sq}(\omega).
\end{equation} 
In analogy to Eq. (\ref{DEGF}), solving this recursive equation for $\widetilde{\cal G}^P$ resums the infinite class of diagrams spanned by the second-order perturbative truncation of the time-ordered self energy $\widetilde\Sigma^{\bigstar}\approx[\widetilde\Sigma^{\bigstar}]^{(P,2)}\equiv\widetilde\Sigma_{P}^{\bigstar}.$ Equation (\ref{intG2}) has the inverse solution
\begin{equation}
[\widetilde{\cal G}_P^{-1}]_{pq}(\omega)=[\widetilde G_{(0)}^{-1}]_{pq}(\omega)-[\widetilde\Sigma^{\bigstar}_P]_{pq}(\omega),
\end{equation} 
which may be written in matrix form as
\begin{equation}
%\begin{split}
\widetilde{\ensuremath{{\mathbfcal G}}}_{P}^{-1}=\left[
\begin{array}{cc}
[\widetilde {\bf G}_{(0)}^{-1}]_\bullet-\widetilde{\bm\Sigma}^{\bigstar}_{P\bullet}&-\widetilde{\bm\Sigma}^{\bigstar}_{P>}\\
-\widetilde{\bm\Sigma}^{\bigstar}_{P\vee}&[\widetilde {\bf G}_{(0)}^{-1}]_\circ-\widetilde{\bm\Sigma}^{\bigstar}_{P\circ}
\end{array}
\right],
%&=
%\left[
%\begin{array}{cc}
%\omega-\varepsilon_p^0/\hbar-\widetilde{\Sigma}_{pp}^{\bigstar P}&-\widetilde{\Sigma}_{pq}^{\bigstar P}\\
%-\widetilde{\Sigma}_{qp}^{\bigstar P}&\omega-\varepsilon_q^0/\hbar-\widetilde{\Sigma}_{qq}^{\bigstar P}
%\end{array}
%\right],
%&=\frac{{\textrm{adj}}\widetilde{\ensuremath{{\mathbfcal G}}}_{P}^{-1}}{\det\widetilde{\ensuremath{{\mathbfcal G}}}_{P}^{-1}}\\
%\left[
%\begin{array}{cc}
%\omega-\varepsilon_p^0/\hbar-\widetilde{\Sigma}_{pp}^{\bigstar P}&-\widetilde{\Sigma}_{pq}^{\bigstar P}\\
%-\widetilde{\Sigma}_{qp}^{\bigstar P}&\omega-\varepsilon_q^0/\hbar-\widetilde{\Sigma}_{qq}^{\bigstar P}
%\end{array}
%\right],
%\end{split}
\end{equation} 
where we have used the fact that the one-body HF Green's function (\ref{LG}) is diagonal in its indices; $[\widetilde {\bf G}_{(0)}^{-1}]_\bullet-\widetilde{\bm\Sigma}^{\bigstar}_{P\bullet}$ and $[\widetilde {\bf G}_{(0)}^{-1}]_\circ-\widetilde{\bm\Sigma}^{\bigstar}_{P\circ}$ represent diagonal particle-particle and hole-hole matrices, while $\widetilde{\bm\Sigma}^{\bigstar}_{P>}$ and $\widetilde{\bm\Sigma}^{\bigstar}_{P\vee}$ represent off-diagonal particle-hole and hole-particle block matrix contributions to $\widetilde{\ensuremath{{\mathbfcal G}}}_{P}^{-1}.$ Formal inversion may be expressed in terms of minors as
\begin{equation}
\widetilde{\ensuremath{{\mathbfcal G}}}^{P}_{pq}=[\det\widetilde{\ensuremath{{\mathbfcal G}}}_{P}^{-1}]^{-1}{(-)^{p+q}{\textrm{minor}}[\widetilde{\ensuremath{{\mathbfcal G}}}_{P}^{-1}]_{qp}}.
\end{equation} 
Alternatively, we make use of the block-matrix inverse
\begin{widetext}
\begin{equation}
\label{invG}
\begin{split}
\widetilde{\ensuremath{{\mathbfcal G}}}^{P}&=
\left[
\begin{array}{cc}
([\widetilde {\bf G}_{(0)}^{-1}]_\bullet-\widetilde{\bm\Sigma}^{\bigstar}_{P\bullet})^{-1}&([\widetilde {\bf G}_{(0)}^{-1}]_\bullet-\widetilde{\bm\Sigma}^{\bigstar}_{P\bullet})^{-1}\widetilde{\bm\Sigma}^{\bigstar}_{P>}([\widetilde {\bf G}_{(0)}^{-1}]_\circ-\widetilde{\bm\Sigma}^{\bigstar}_{P\circ})^{-1}\\
([\widetilde {\bf G}_{(0)}^{-1}]_\circ-\widetilde{\bm\Sigma}^{\bigstar}_{P\circ})^{-1}\widetilde{\bm\Sigma}^{\bigstar}_{P\vee}([\widetilde {\bf G}_{(0)}^{-1}]_\bullet-\widetilde{\bm\Sigma}^{\bigstar}_{P\bullet})^{-1}&([\widetilde {\bf G}_{(0)}^{-1}]_\circ-\widetilde{\bm\Sigma}^{\bigstar}_{P\circ})^{-1}
\end{array}
\right]
%\frac{1}{D}\left[
%\begin{array}{cc}
%[\widetilde {\bf G}_{(0)}^{-1}]_\circ-\widetilde{\bm\Sigma}^{\bigstar}_{P\circ}(\omega)&\widetilde{\bm\Sigma}^{\bigstar}_{P>}(\omega)\\
%\widetilde{\bm\Sigma}^{\bigstar}_{P\vee}(\omega)&[\widetilde {\bf G}_{(0)}^{-1}]_\bullet-\widetilde{\bm\Sigma}^{\bigstar}_{P\bullet}(\omega)
%\end{array}
%\right]\\
%&=
%D^{-1}\left[
%\begin{array}{cc}
%\omega-\varepsilon_i^0/\hbar-\widetilde{\Sigma}_{ii}^{\bigstar P}&\widetilde{\Sigma}_{fi}^{\bigstar P}\\
%\widetilde{\Sigma}_{if}^{\bigstar P}&\omega-\varepsilon_f^0/\hbar-\widetilde{\Sigma}_{ff}^{\bigstar P}
%\end{array}
%\right]
\end{split}
\end{equation}
\end{widetext}
together with the weak-coupling approximation where $[\widetilde\Sigma^{\bigstar}_P]_{pq},[\widetilde\Sigma^{\bigstar}_P]_{qp}\ll[\widetilde\Sigma^{\bigstar}_P]_{pp},[\widetilde\Sigma^{\bigstar}_P]_{qq}.$ Here, the infinitesimals $\pm i0^+$ are omitted from $G^{(0)}$ as the self energy (\ref{ISE}) is complex valued and has an imaginary component that changes sign with respect to the chemical potential according to ${\textrm{Im}}{\Sigma}^{\bigstar}_P(\omega)>0$ when $\omega<\mu/\hbar$ and ${\textrm{Im}}{\Sigma}^{\bigstar}_P(\omega)<0$ when $\omega>\mu/\hbar.$ Like $\widetilde G^{(0)}_{pq},$ $\widetilde {\cal G}^{P}_{pq}$ in Eq. (\ref{invG}) accounts for the propagation of an extra electron from the single-particle state $q$ to $p;$ however, unlike the noninteracting Green's function where $p=q,$ $\widetilde {\cal G}^{P}_{pq}$ also permits electrons to scatter out of $q$ and into the single-particle state $p\neq q.$ This latter process is described through the off-diagonal components
\begin{equation}
\label{G2M}
\begin{split}
&\widetilde {\cal G}^{P}_{ib}(\omega)\\
&=\sum_{ja}\frac{\delta_{ij}}{\omega-\varepsilon_{i}^0/\hbar-[\widetilde{\Sigma}^{\bigstar}_P]_{ii}(\omega)}[\widetilde{\Sigma}^{\bigstar}_P]_{ja}(\omega)\frac{\delta_{ab}}{\omega-\varepsilon_{b}^0/\hbar-[\widetilde{\Sigma}^{\bigstar}_P]_{bb}(\omega)}\\
&\widetilde {\cal G}^{P}_{aj}(\omega)\\
&=\sum_{ib}\frac{\delta_{ab}}{\omega-\varepsilon_{a}^0/\hbar-[\widetilde{\Sigma}^{\bigstar}_P]_{aa}(\omega)}[\widetilde{\Sigma}^{\bigstar}_P]_{bi}(\omega)\frac{\delta_{ij}}{\omega-\varepsilon_{j}^0/\hbar-[\widetilde{\Sigma}^{\bigstar}_P]_{jj}(\omega)},
\end{split}
\end{equation}
which are here derived in the weak-coupling limit.

This interacting molecular-electronic one-body Green's function $\widetilde {\cal G}^{P}$ has the space- and frequency-dependent form
\begin{equation}
\label{dysonG}
%\begin{split}
\widetilde {\cal G}^{P}({\bf x},{\bf x}';\omega)=\sum_{pq}\chi_{p}({\bf x})\widetilde {\cal G}^{P}_{pq}(\omega)\chi_{q}^*({\bf x}')
%&={\widetilde G}^{(0)}({\bf x},{\bf x}';\omega)\ \ \ \ \ +{\tint}d^3yd^3y'{\widetilde G}^{(0)}({\bf x},{\bf y};\omega)\widetilde{\Sigma}^{\bigstar}_P({\bf y},{\bf y}';\omega)\widetilde {\cal G}^{P}({\bf y}',{\bf x}';\omega)\\
%&=\sum_{pq}\chi_{p}({\bf x})\widetilde {\cal G}^{P}_{pq}(\omega)\chi_{q}^*({\bf x}')
%\end{split}
\end{equation}
with interacting Dyson orbitals $\chi_{q}({\bf x})=\sum_{r}\phi_r({\bf x})U_{rq}$ satisfying the nonlinear integro-differential equation
\begin{widetext}
\begin{equation}
\label{phiint1}
%\begin{split}
[h_0({\bf x})+U_0({\bf x})]\chi_{q}({\bf x})+\sum_k[\langle k|V|k\rangle({\bf x})\chi_{q}({\bf x})-\langle k|V|{q}\rangle({\bf x})\chi_k({\bf x})]+{\tint}d^3x'\hbar\widetilde\Sigma^{\bigstar}_{P}({\bf x},{\bf x}';\omega)\chi_{q}({\bf x}')=[\varepsilon^0_{q}+\hbar[\widetilde\Sigma^{\bigstar}_P]_{qq}(\omega)]\chi_{q}({\bf x})
%&0=[h_0({\bf x})+U_0({\bf x})-\varepsilon^0_q-\hbar\widetilde\Sigma^{\bigstar P}_{qq}(\omega)]\chi_q({\bf x})+\sum_i[\langle i|V|i\rangle({\bf x})\chi_q({\bf x})-\langle i|V|q\rangle({\bf x})\chi_i({\bf x})]+{\tint}d^3x'\hbar\widetilde\Sigma^{\bigstar}_{P}({\bf x},{\bf x}';\omega)\chi_q({\bf x}')\\
%&+\sum_{pi}\frac{\widetilde\Sigma^{\bigstar P}_{pq}(\omega)}{\omega-\varepsilon^0_p/\hbar-\widetilde\Sigma^{\bigstar P}_{pp}(\omega)}\Bigl\{[h_0({\bf x})+U_0({\bf x})-\hbar\omega]\chi_p({\bf x})+[\langle i|V|i\rangle({\bf x})\chi_p({\bf x})-\langle i|V|p\rangle({\bf x})\chi_i({\bf x})]+{\tint}d^3x'\hbar\widetilde\Sigma^{\bigstar}_{P}({\bf x},{\bf x}';\omega)\chi_p({\bf x}')\Bigr\}
%\end{split}
\end{equation}
\end{widetext}
expressed in terms of the nonlocal frequency-dependent (energy-dependent) effective potential $\widetilde{\Sigma}^{\bigstar}_{P}({\bf x},{\bf x}';\omega)=\sum_{pq}\chi_{p}({\bf x})[\widetilde{\Sigma}^{\bigstar}_P]_{pq}(\omega)\chi_{q}^*({\bf x}').$ It is derived by applying the operator 
\begin{equation}
\begin{split}
L_{\textrm{HF}}({\bf x}&,{\bf x}';\omega)=[\hbar\omega-\{h_0({\bf x})+U_0({\bf x})\}]\delta({\bf x}-{\bf x}')\\
&-\sum_k[\langle k|V|k\rangle({\bf x})\delta({\bf x}-{\bf x}')-\chi_k^*({\bf x}')V({\bf x},{\bf x}')\chi_k({\bf x})],
\end{split}
\end{equation}
which is defined in terms of $\chi_k,$ to $\widetilde {\cal G}^{P}$ in Eq. (\ref{dysonG}) and then projecting against $\chi_q;$ $L_{\textrm{HF}}$ can equivalently be written in terms of the HF orbitals $\phi_k$ where it satisfies $\int d^3yL_{\textrm{HF}}({\bf x},{\bf y};\omega)\widetilde G^{(0)}({\bf y},{\bf x}';\omega)=\hbar\delta({\bf x}-{\bf x}').$ Equation (\ref{phiint1}) incorporates the image interaction associated with the molecule-induced excitations of the particle into the molecular HF mean-field equations (\ref{HFE}) through the irreducible self energy $\hbar\Sigma^{\bigstar}_{P}$ in Eq. (\ref{ISE}). These interacting orbitals and orbital energies, which are solutions of Eq. (\ref{phiint1}), must reach self consistency with $\hbar[\widetilde\Sigma^{\bigstar}_P]_{pq}$ defined in Eq. (\ref{RPAISE}) as the self energy both determines and is determined by the new interacting orbitals. In this way, consistent solutions of the interacting molecule-particle system may be obtained.

%In the weak-coupling limit, these Dyson orbitals are solutions of the approximate equation
%\begin{equation}
%\label{phiint2}
%\begin{split}
%0&=[h_0({\bf x})+U_0({\bf x})-\varepsilon^0_q-\hbar\widetilde\Sigma^{\bigstar P}_{qq}(\omega)]\chi_q({\bf x})\\
%&\ \ \ +\sum_i[\langle i|V|i\rangle({\bf x})\chi_q({\bf x})-\langle i|V|q\rangle({\bf x})\chi_i({\bf x})]\\
%&\ \ \ +{\tint}d^3x'\hbar\widetilde\Sigma^{\bigstar}_{P}({\bf x},{\bf x}';\omega)\chi_q({\bf x}').
%\end{split}
%\end{equation}

\subsubsection{\label{eep}Interaction of coupled molecule-particle system with an external electric field}
Now that we have derived an expression for the molecular one-body Green's function interacting with its self-induced electronic density fluctuations in a nearby metal particle ({\it i.e.}, its image), we are ready to build in the perturbing effects of an external classical electric field 
\begin{equation}
{\bf E}_0(t)={\bf E}^{(+)}_0e^{-i\omega_{\bf k}t}+{\bf E}^{(-)}_0e^{i\omega_{\bf k}t}
\end{equation}
with frequency $\omega_{\bf k}$ upon the combined and coupled system. The field interacts directly with the molecule and, additionally, indirectly by inducing small-amplitude collective excitations of conduction electrons in the particle described by $n_E.$ This latter interaction is described in Ref. \cite{nitzan1980} as the {\it local-field effect.} Following Eq. (\ref{density}), the particle's electronic density can be decomposed as
\begin{equation}
\begin{split}
n({\bf x},t)&=n_0({\bf x})+n_E({\bf x},t)\\
&=n_0({\bf x})+n^{(+)}_E({\bf x},t)+n^{(-)}_E({\bf x},t)
\end{split}
\end{equation}
in the linear response limit of the external field, where the labels $(+),(-)$ refer to excitations set up by the incoming and outgoing components ${\bf E}^{(+)}_0$ and ${\bf E}^{(-)}_0$ respectively of ${\bf E}_0.$ Despite our classical treatment of the external electric field, we continue to speak of photons and will, when justifiable and appropriately pointed out, need to make an {\it ad hoc} adjustment of photon occupation number as a result of our classical-field ansatz. Of course, we could have treated the external field quantum mechanically and introduced the appropriate photon Green's functions needed to carry its dynamics. However, it is our aim to keep this presentation as clear as possible and elucidate only the lowest-order processes governing SERS, which, save a small error in photon occupation number, is describable without resorting to field quantization.

Multipole expansion of the electron-field and electron-plasmon interaction Hamiltonian $\hat F_{\textrm{int}},$ given in Eq. (\ref{Fint}), results in the following expression
\begin{equation}
\begin{split}
F^{\textrm{int}}_{pq}(t)&=-\langle p|-e{\bf x}|q\rangle\cdot{\bf E}_0(t)+\langle p|{\tint}d^3xn_E({\bf x},t)V({\bf x},{\bf x}')|q\rangle\\
&\approx-\langle p|-e{\bf x}|q\rangle\cdot\Bigl\{{\bf E}_0(t)+\frac{3\hat{\bf r}[{\bf p}_E(t)\cdot\hat{\bf r}]-{\bf p}_E(t)}{r^3}\Bigr\}
\end{split}
\end{equation}
for its matrix elements at dipole order. The second term in brackets is the classical electric dipole field ${\bf E}_P(t)={\bm\Lambda}\cdot{\bf p}_E(t)={\bm\Lambda}\cdot\int(-e{\bf x})n_E({\bf x},t)d^3x$ of the metal particle's dipole plasmon as induced by ${\bf E}_0.$ Like Eq. (\ref{UM}), it contributes the effective plasmon potential
\begin{equation}
\label{UE}
\begin{split}
U_{\textrm{int}}({\bf x},t)\to U_E({\bf x},t)&\approx -(-e{\bf x})\cdot{\bm\Lambda}\cdot{\bf p}_E(t)\\
&=-(-e{\bf x})\cdot{\bf E}_P(t),
\end{split}
\end{equation}
where, like $n_E$ and ${\bf p}_E,$ ${\bf E}_P$ is an abbreviation for ${\bf E}^{(+)}_P+{\bf E}^{(-)}_P.$ Since we choose to specialize to the dipole approximation for simplicity, no other multipole contributions will be considered. Therefore, perturbations to the molecular-electronic system from ${\bf E}_0$ arise directly and indirectly through
\begin{equation}
\label{fep1}
\begin{split}
\hat F_{\textrm{int}}&(t)\to\hat F_E(t)=-\hat{\bf d}(t)\cdot[{\bf E}_0(t)+{\bf E}_P(t)]\\
&=-\sum_{pq}\langle p|-e{\bf x}|q\rangle N\{\hat C^\dagger_p(t)\hat C_q(t)\}\cdot[{\bf E}_0(t)+{\bf E}_P(t)],
\end{split}
\end{equation} 
where the new interacting molecular electron operators $\hat C_p(t)=\sum_qU_{pq}\hat c_q(t)$ are related to the old noninteracting operators $\hat c_p(t)$ by the same unitary transformation $U$ that affected the molecular orbitals.

As before, Dyson's expansion provides a systematic way to build in the interaction effects of $\hat F_E$ into the molecular one-body Green's function. In contrast to the previous application of the Dyson expansion where the interaction effects of $\hat F_P$ were included on top of the noninteracting HF one-body Green's function $G^{(0)},$ here we build the interaction effects of $\hat F_E$ upon the interacting one-body Green's function ${\cal G}^{P}_{pq}(t,t')=\langle\Phi^N_{\textrm{HF}}|T\{\hat C_p(t)\hat C_q^\dagger(t')\}|\Phi^N_{\textrm{HF}}\rangle$ that describes the coupling between molecular electrons and their image as mediated by the conduction electrons of a metallic particle. We see that
\begin{widetext}
\begin{equation}
\label{linkE}
\begin{split}
&i{\cal G}^{\textrm{int}}_{pq}(t,t')\\
&=\sum_{n=0}^\infty\frac{(-i/\hbar)^n}{n!}{\tint}dt_1\cdots dt_n\langle\Phi^N_{\textrm{HF}}|T\{-\hat{\bf d}(t_1)\cdot[{\bf E}_0(t_1)+{\bf E}_P(t_1)]\cdots-\hat{\bf d}(t_n)\cdot[{\bf E}_0(t_n)+{\bf E}_P(t_n)]\hat C_p(t)\hat C_q^\dagger(t')\}|\Phi^N_{\textrm{HF}}\rangle_C\\
&=i{\cal G}^{P}_{pq}(t,t')+\frac{i}{\hbar}\sum_{rs}{\tint}dt_1{\cal G}^{P}_{pr}(t,t_1)\langle r|e{\bf x}|s\rangle {\cal G}^{P}_{sq}(t_1,t')\cdot[{\bf E}_0(t_1)+{\bf E}_P(t_1)]\\
&\ \ \ +\frac{i}{\hbar^2}\sum_{rstu}{\tint}dt_1dt_2 {\cal G}^{P}_{pr}(t,t_1)\langle r|-ex^\xi|t\rangle {\cal G}^{P}_{tu}(t_1,t_2)\langle u|-ex^{\prime\sigma}|s\rangle {\cal G}^{P}_{sq}(t_2,t')\big\{E^\xi_0(t_1)E^\sigma_0(t_2)+E^\xi_0(t_1)E^\sigma_P(t_2)\\
&\hspace{3.5cm}+E^\xi_P(t_1)E^\sigma_0(t_2)+E^\xi_P(t_1)E^\sigma_P(t_2)\big\}+\cdots,
\end{split}
\end{equation} 
\end{widetext}
where, as was discussed previously in the context of $\Pi_P,$ the time-ordering affects the classical fields through
\begin{equation}
\begin{split}
&E^\xi_0(t_1)E^\sigma_0(t_2)=T\big\{E^\xi_0(t_1)E^\sigma_0(t_2)\big\}\\
&=\theta(t_1-t_2)E^\xi_0(t_1)E^\sigma_0(t_2)+\theta(t_2-t_1)E^\sigma_0(t_2)E^\xi_0(t_1).
\end{split}
\end{equation} 
Similar expressions can be written for the remaining three terms in curly brackets above. Truncating the perturbation series in Eq. (\ref{linkE}) at second order in $\hat F_E$ results in the second-order perturbative approximation to the interacting one-body Green's function
\begin{widetext}
\begin{equation}
\label{linkE2}
\begin{split}
{\cal G}^{\textrm{int}(2)}_{pq}(t,t')&={\cal G}^{P}_{pq}(t,t')+\sum_{rs}{\tint}dt_1dt_1'{\cal G}^{P}_{pr}(t,t_1)\Sigma_{rs}^{\bigstar(E,2)}(t_1,t_1'){\cal G}^{P}_{sq}(t_1,t')\\
&={\cal G}^{P}_{pq}(t,t')+\sum_{rs}{\tint}dt_1dt_2 {\cal G}^{P}_{pr}(t,t_1)\Bigl[\frac{1}{\hbar^2}\sum_{tu}\langle r|-ex^\xi|t\rangle {\cal G}^{P}_{tu}(t_1,t_2)\langle u|-ex^{\prime\sigma}|s\rangle\big\{E^\xi_0(t_1)E^\sigma_0(t_2)\\
&\hspace{3.5cm}+E^\xi_0(t_1)E^\sigma_P(t_2)+E^\xi_P(t_1)E^\sigma_0(t_2)+E^\xi_P(t_1)E^\sigma_P(t_2)\big\}\Bigr] {\cal G}^{P}_{sq}(t_2,t'),
\end{split}
\end{equation} 
\end{widetext}
where we have omitted the first-order perturbative correction as it describes only the stimulated absorption and emission of ${\bf E}_0$ and ${\bf E}_P$ which are of no importance in the following (and can be renormalized away).

%!!!!!!!!!!!!!!!!!!!!!!!!!!!!!!!!!!!!!!!!!!!!!!!!!!!!!!!!!!!
%!!!!!!!!!!!!!!!!!!!!!!!!!!!!!!!!!!!!!!!!!!!!!!!!!!!!!!!!!!!
\begin{figure*}
%\psfrag{p}[][]{{\Huge $p$}}
%\psfrag{q}[][]{{\Huge $q$}}
%\psfrag{E01}[][]{{\Huge $\widetilde{\bf E}_0^{(+)}$}}
%\psfrag{EP1}[][]{{\Huge $\widetilde{\bf E}_{0P}^{(+)}$}}
%\psfrag{E02}[][]{{\Huge $\widetilde{\bf E}_0^{(-)}$}}
%\psfrag{EP2}[][]{{\Huge $\widetilde{\bf E}_{0P}^{(-)}$}}
%\psfrag{a}[][]{{\Huge (a.)}}
%\psfrag{b}[][]{{\Huge (b.)}}
%\psfrag{c}[][]{{\Huge (c.)}}
%\psfrag{d}[][]{{\Huge (d.)}}
%\psfrag{G}[][]{{\Huge $\widetilde{\cal G}^{\textrm{int}(2)}_{pq}(\omega)-\widetilde{\cal G}^P_{pq}(\omega)$}}
\begin{center}
%\rotatebox{0}{\resizebox{!}{3.9cm}{\includegraphics{raman1fig}}}
\rotatebox{0}{\resizebox{!}{4.3cm}{\includegraphics{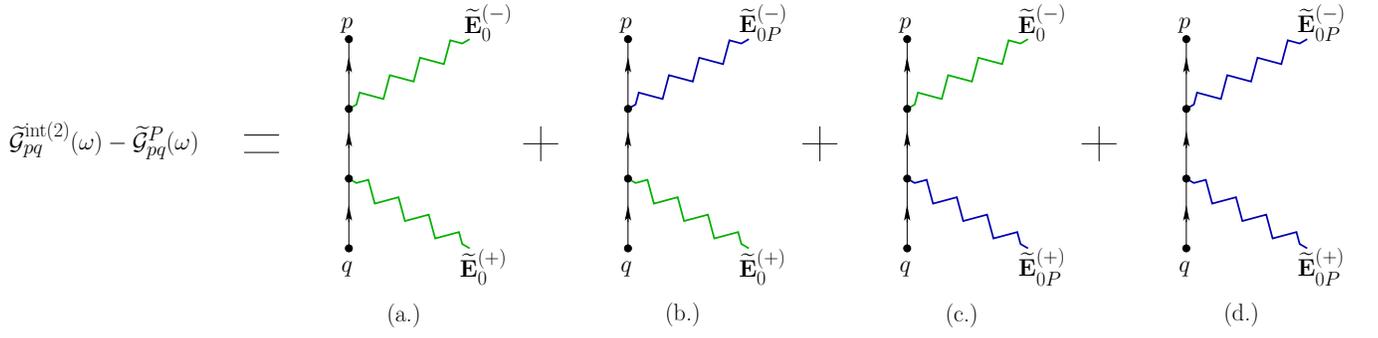}}}
\caption{\label{ram0} (Color online) Enhanced Raman-scattering Feynman diagrams occurring at second order in perturbation theory in the interaction $\hat F_E=-\hat{\bf d}\cdot[{\bf E}_0+{\bf E}_P].$ Diagram (a.) is analogous to normal Raman scattering of the field $\widetilde{\bf E}_0^{(\pm)}.$ Diagrams (b.) and (c.) represent a particle-mediated Raman-scattering process where either the Raman scattered field [diagram (b.)] or the incident field [diagram (c.)] is scattered by the particle. Diagram (d.) represents the process where both the incident and Raman-scattered fields are mediated by the particle. This particle-mediated field $\widetilde{\bf E}_{0P}^{(\pm)}={\bm\Lambda}\cdot\widetilde{\bm\alpha}_P\cdot\widetilde{\bf E}^{(\pm)}_0$ is defined in terms of the particle's polarizability $\widetilde{\bm\alpha}_P$ (\ref{alphaP}). While retained in the formalism [see, {\it e.g.}, Eq. (\ref{ISEE})], crossed diagrams are not drawn for simplicity in presentation. However, all two-photon absorption and two-photon emission processes are omitted here and in the theory as are all first-order perturbative contributions. Note that all molecular-electronic propagators are interacting one-body Green's functions $\widetilde{\cal G}^P$ defined in Eq. (\ref{invG}).}
\end{center}
\end{figure*}
%!!!!!!!!!!!!!!!!!!!!!!!!!!!!!!!!!!!!!!!!!!!!!!!!!!!!!!!!!!!
%!!!!!!!!!!!!!!!!!!!!!!!!!!!!!!!!!!!!!!!!!!!!!!!!!!!!!!!!!!!
Indeed it is not until second order in the external field perturbation that Raman scattering from the molecular system can be described. Other competing processes occur at second order as well, such as those of two photon-absorption or two-photon emission. However, we are not interested in describing these events and, consequently, prune away all terms in Eq. (\ref{linkE2}) not related to Raman scattering. Due to the presence of both ${\bf E}_0$ and ${\bf E}_P,$ several types of Raman processes are present in the remaining expression. Note that the electric field ${\bf E}_P$ stems from collective excitation of conduction electrons in the metal particle induced either by the external field ({\it i.e.}, the local-field effect) or by the Raman-scattered field of the molecule. In either case, it may be written in Fourier space as
\begin{equation}
\label{below}
\begin{split}
\widetilde{\bf E}_P^{(\pm)}(\omega)&={\bm\Lambda}\cdot\widetilde{\bf p}^{(\pm)}_E(\omega)\\
%&={\bm\Lambda}\cdot[\widetilde{\bf p}^{(\pm)}(\omega)-{\bf p}_0]\\
%&={\bm\Lambda}\cdot\widetilde{\bm\alpha}_{\textrm{RPA}}(\omega)\cdot\widetilde{\bf E}^{(\pm)}_0(\omega)\\
&=\{{\bm\Lambda}\cdot\widetilde{\bm\alpha}_P(\omega)\cdot\widetilde{\bf E}^{(\pm)}_0\}2\pi\delta(\omega\pm\omega_{{\bf k},{\bf k}'})\\
&=\widetilde{\bf E}^{(\pm)}_{0P}(\omega)2\pi\delta(\omega\pm\omega_{{\bf k},{\bf k}'})
\end{split}
\end{equation} 
in terms of the linear polarizability defined in Eq. (\ref{alphaP}), where the wave vector $\bf k$ is associated with the incoming field labeled by $(+)$ while the wave vector ${\bf k}'$ is associated with the outgoing field labeled by $(-);$ they satisfy $[\widetilde{\bf E}^{(+)}_0]^*=\widetilde{\bf E}^{(-)}_0.$

Together with this expression for $\widetilde{\bf E}_P^{(\pm)},$ the interacting one-body Green's function in Eq. (\ref{linkE2}) enjoys the Fourier decomposition
\begin{equation}
\label{DEGFE}
\widetilde{\cal G}^{\textrm{int}(2)}_{pq}(\omega)=\widetilde{\cal G}^P_{pq}(\omega)+\sum_{rs}\widetilde{\cal G}^P_{pr}(\omega)\widetilde\Sigma_{rs}^{\bigstar(E,2)}(\omega)\widetilde{\cal G}^P_{sq}(\omega)
\end{equation} 
with second-order irreducible self energy for Raman scattering
\begin{equation}
\label{ISEE}
\begin{split}
\hbar&\widetilde\Sigma_{pq}^{\bigstar(E,2)}(\omega)\approx\frac{1}{\hbar}\sum_{rs}\langle p|-ex^\xi|r\rangle \widetilde{\cal G}^{P}_{rs}(\omega)\langle s|-ex^{\prime\sigma}|q\rangle\\
&\ \times\left\{
\begin{array}{l}
\widetilde E^{(-)}_{0\xi}\widetilde E^{(+)}_{0\sigma}+\widetilde E^{(+)}_{0\xi}\widetilde E^{(-)}_{0\sigma}\vspace{0.1cm}\\
+\widetilde E^{(-)}_{0P\xi}(-\omega_{{\bf k}'})\widetilde E^{(+)}_{0\sigma}+\widetilde E^{(+)}_{0P\xi}(\omega_{{\bf k}})\widetilde E^{(-)}_{0\sigma}\vspace{0.1cm}\\
+\widetilde E^{(-)}_{0\xi}\widetilde E^{(+)}_{0P\sigma}(\omega_{\bf k})+\widetilde E^{(+)}_{0\xi}\widetilde E^{(-)}_{0P\sigma}(-\omega_{{\bf k}'})\vspace{0.1cm}\\
+\widetilde E^{(-)}_{0P\xi}(-\omega_{{\bf k}'})\widetilde E^{(+)}_{0P\sigma}(\omega_{\bf k})+\widetilde E^{(+)}_{0P\xi}(\omega_{{\bf k}})\widetilde E^{(-)}_{0P\sigma}(-\omega_{{\bf k}'}),
\end{array}
\right.
%\big\{\widetilde E^{(-)}_{0\xi}\widetilde E^{(+)}_{0\sigma}+\widetilde E^{(+)}_{0\xi}\widetilde E^{(-)}_{0\sigma}\\
%&\hspace{.5cm}+\widetilde E^{(-)}_{0P\xi}(-\omega_{{\bf k}'})\widetilde E^{(+)}_{0\sigma}+\widetilde E^{(+)}_{0P\xi}(\omega_{{\bf k}})\widetilde E^{(-)}_{0\sigma}\\
%&\hspace{.5cm}+\widetilde E^{(-)}_{0\xi}\widetilde E^{(+)}_{0P\sigma}(\omega_{\bf k})+\widetilde E^{(+)}_{0\xi}\widetilde E^{(-)}_{0P\sigma}(-\omega_{{\bf k}'})\\
%&\hspace{.5cm}+\widetilde E^{(-)}_{0P\xi}(-\omega_{{\bf k}'})\widetilde E^{(+)}_{0P\sigma}(\omega_{\bf k})+\widetilde E^{(+)}_{0P\xi}(\omega_{{\bf k}})\widetilde E^{(-)}_{0P\sigma}(-\omega_{{\bf k}'})\big\},\\
%&\frac{1}{\hbar}\sum_{rs}\langle p|ex^\xi|r\rangle \widetilde{\cal G}^{P}_{rs}(\omega)\langle s|ex^{\prime\sigma}|q\rangle\big\{\widetilde E^{(-)}_{0\xi}\widetilde E^{(+)}_{0\sigma}+\widetilde E^{(-)}_{0\xi}\widetilde E^{(+)}_{0P\sigma}(\omega_{\bf k})\\
%&\hspace{1.5cm}+\widetilde E^{(-)}_{0P\xi}(-\omega_{{\bf k}'})\widetilde E^{(+)}_{0\sigma}+\widetilde E^{(-)}_{0P\xi}(-\omega_{{\bf k}'})\widetilde E^{(+)}_{0P\sigma}(\omega_{\bf k})\big\},
\end{split}
\end{equation}
where the energies of the intermediate electronically excited molecular states $r$ and $s$ both include the additional energy $\hbar\omega_{\bf k}$ ($-\hbar\omega_{{\bf k}'}$) of a single absorbed (emitted) photon. By replacing $\widetilde{\cal G}^P$ by $\widetilde G^{(0)},$ the first pair of terms in brackets represents ordinary Raman scattering from the molecule in the absence of the particle. Terms three and four are associated with the mixed event in which the incident field directly interacts with the molecule while the molecular Raman-scattered field scatters off of the particle before detection; terms five and six represent the opposite time ordering of terms three and four. The last pair of terms are associated with the scattering event where the incident field is first scattered by the particle. This enhanced field interacts with the molecule, which subsequently Raman scatters the radiation back to the particle. In the final step, the particle rebroadcasts the molecular Raman field to the detector. These processes are all summarized in Fig. \ref{ram0}. While they are present in the formalism, crossed events where a photon is first scattered and, later, a second photon is absorbed are omitted from the figure. As we are interested only in Raman scattering, two-photon absorption/emission processes, which would involve terms like $\widetilde E^{(+)}_{0\xi}\widetilde E^{(+)}_{0\sigma}$ or $\widetilde E^{(-)}_{0\xi}\widetilde E^{(-)}_{0\sigma}$ are not considered. Note that additional diagrams would be present had we chosen to quantize the electric field. Henceforth, for simplicity of notation we drop the label 2 so that $\widetilde\Sigma_{E}^{\bigstar}\equiv\widetilde\Sigma_{(E,2)}^{\bigstar}.$

\subsubsection{\label{RPAsec}Random-phase approximation}
Until this point we have assumed that the collective excitations of conduction electrons in the metal particle can be represented entirely by their classical density $n.$ When a molecular-electronic system is brought into the vicinity of a particle it induces density fluctuations in the metal that are, to first order, describable by $n_M=n-n_0.$ For the purpose of demonstrating certain properties of our formalism, we here invoke the further approximation that the conduction electrons of the spatially anisotropic metal particle are well-described as a homogeneous electron gas (or electron plasma) in the high-density limit. This approximation becomes appropriate when the spatial dimensions of the particle are larger than the mean-free path of its conduction electrons. For Au and Ag, which are typical SERS substrates, the mean-free path of conduction electrons is approximately 40 and 50 nm respectively. This justifies the replacement of the particle's polarization propagator by the random-phase approximation (RPA) result \cite{Bohm52,McLachlan64}, {\it i.e.}, $\Pi_{P}\to\Pi{_{\textrm{RPA}}}=\Pi_{P}(1-F_P\Pi_{P})^{-1}=\Pi_{P}\kappa^{-1},$ where $\kappa=1-(\Omega_{0}/\omega)^2$ is the generalized dielectric function of the particle in the RPA. Similarly, the polarizability of the metal particle takes on the RPA form
\begin{equation}
\label{RPAa}
\widetilde\alpha_{P}^{\xi\sigma}(\omega)\to\widetilde\alpha_{\textrm{RPA}}^{\xi\sigma}(\omega)=\frac{-e^2/m}{\omega^2-\Omega_{0}^2}\delta^{\xi\sigma},
\end{equation}
where $\Omega_{0}=\sqrt{4\pi e^2n_F/m}$ is the bulk plasma frequency and $n_F=k_F^3/3\pi^2$ is the density of the free electron gas with Fermi wave vector $k_F=\sqrt{2m\varepsilon_F/\hbar^2}.$ It is, of course, a severe approximation to assume that the polarization propagator of an arbitrarily sized and shaped particle will have the RPA form. (Extension to a damped Drude or Lorentz oscillator model with several resonant frequencies would be straightforward.) Rather the dynamics of the metal particle's conduction electrons and their induced dipole moments ${\bf p}_M$ should be solved for explicitly. While this ultimately is our desire and is the subject of Sec. \ref{PRP1} below, such a task requires tremendous numerical effort and, at this stage, it is only our intent to lay out the basic working equations of our model and to highlight some of its general results and salient features. Choosing, for this purpose, to temporarily make a detour and invoke the RPA provides a physically reasonable analytical model of the particle's response that is sufficiently rich to allow us to do so without having to explicitly compute the electronic dynamics of the metal particle.

\begin{figure}[t]
%\psfrag{p}[][]{{\LARGE $q$}}
%\psfrag{q}[][]{{\LARGE $p$}}
%\psfrag{S}[][]{{\huge $\widetilde\Pi_{\textrm{RPA}}(\omega)$}}
%\psfrag{G1}[][]{{\huge${\widetilde{\cal G}}^{\textrm{\Large{\it P}}}_{\textrm{\Large{$pq$}}}(\omega)$}}
\begin{center}
%\rotatebox{0}{\resizebox{!}{4cm}{\includegraphics{APP1_new}}}
\rotatebox{0}{\resizebox{!}{4cm}{\includegraphics{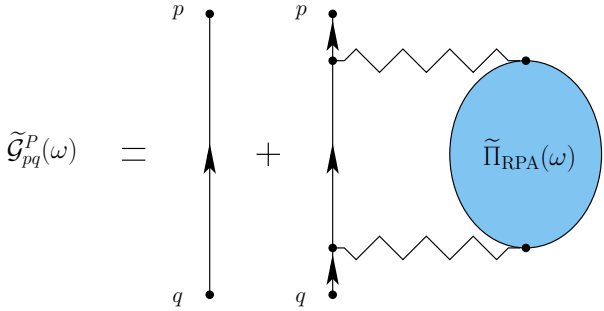}}}
\caption{\label{G} (Color online) Second-order (second Born approximation beyond HF) one-body Green's function ${\widetilde{\cal G}}^{P}$ accounting for the interaction of molecular electrons with collective excitations of conduction electrons in a nearby metal particle described by the RPA polarization propagator $\widetilde\Pi_{\textrm{RPA}}.$ Here, the molecular-electronic propagators are one-body HF Green's functions ${\widetilde G}^{(0)}.$} 
\end{center}
\end{figure}

The self energy $\hbar\widetilde{\Sigma}^{\bigstar}_P,$ accounting for the particle-mediated interaction of molecular electrons with their own image, is hereafter approximated in the RPA (\ref{RPAa}) as
\begin{widetext}
\begin{equation}
\label{RPAISE}
\begin{split}
\hbar[\widetilde\Sigma^{\bigstar}_P]_{pq}(\omega)&\approx\sum_{rs}\langle p|-e{\bf x}|r\rangle\cdot{\bm\Lambda}\cdot{\tint}\frac{d\omega'}{2\pi i}\widetilde{\bm\alpha}_{\textrm{RPA}}(\omega')\widetilde G^{(0)}_{rs}(\omega+\omega')\cdot{\bm\Lambda}\cdot\langle s|-e{\bf x}'|q\rangle\\
%&=\frac{e^2}{2m\Omega_{0}r^3}\sum_{s}\langle p|e{\bf x}|s\rangle\cdot{\bm\Lambda}\cdot\langle s|e{\bf x}'|q\rangle\Big[\frac{1-\rho_s^0}{\omega+i0^+-\varepsilon^0_s/\hbar-\Omega_{0}}+\frac{\rho_s^0}{\omega-i0^+-\varepsilon^0_s/\hbar+\Omega_{0}}\Big]\\
&=\frac{e^2}{2m\Omega_{0}r^3}\sum_{s}\langle p|-e{\bf x}|s\rangle\cdot{\bm\Lambda}\cdot\langle s|-e{\bf x}'|q\rangle\Big[\frac{\rho_s^0}{\varepsilon^0_s/\hbar-\Omega_{0}-\omega+i0^+}+\frac{1-\rho_s^0}{\varepsilon^0_s/\hbar+\Omega_{0}-\omega-i0^+}\Big].
\end{split}
\end{equation} 
\end{widetext} 
Here, the near-idempotency of $\Lambda^{\eta\beta}=-r^3\Lambda^{\eta\sigma}\Lambda^{\sigma\beta}$ has been exploited to simplify the expression. The diagrammatic representation of $\widetilde {\cal G}^{P}_{pq}$ is displayed in Fig. \ref{G}. Note that the sum runs over the set of all single-particle states which depend parametrically upon the $M$ nuclear coordinates.
%!!!!!!!!!!!!!!!!!!!!!!!!!!!!!!!!!!!!!!!!!!!!!!!!!!!!!!!!!!!
%!!!!!!!!!!!!!!!!!!!!!!!!!!!!!!!!!!!!!!!!!!!!!!!!!!!!!!!!!!!
%!!!!!!!!!!!!!!!!!!!!!!!!!!!!!!!!!!!!!!!!!!!!!!!!!!!!!!!!!!!
%!!!!!!!!!!!!!!!!!!!!!!!!!!!!!!!!!!!!!!!!!!!!!!!!!!!!!!!!!!!

%\begin{equation}
%\label{G2M}
%\begin{split}
%&\widetilde {\cal G}^{P}_{pq}(\omega)=\frac{\delta_{pq}}{\omega-\varepsilon_{p}^0/\hbar-[\widetilde{\Sigma}^{\bigstar}_P]_{pp}(\omega)}\\
%&+\sum_{rs}\frac{\delta_{pr}}{\omega-\varepsilon_{p}^0/\hbar-[\widetilde{\Sigma}^{\bigstar}_P]_{pp}(\omega)}[\widetilde{\Sigma}^{\bigstar}_P]_{rs}(\omega)\frac{\delta_{sq}}{\omega-\varepsilon_{q}^0/\hbar-[\widetilde{\Sigma}^{\bigstar}_P]_{qq}(\omega)},
%\end{split}
%\end{equation} 

The complex-valued and frequency-dependent irreducible self energy $\hbar\widetilde\Sigma^{\bigstar}_{P}$ may be decomposed into real and imaginary components as 
\begin{equation}
[\widetilde\Sigma^{\bigstar}_{R}]^P(\hbar\omega+i0^+)=\Delta^P(\hbar\omega)-(i/2)\Gamma^P(\hbar\omega).
\end{equation} 
By appealing to the identity $(x\pm i0^+)^{-1}=P(x^{-1})\mp\pi i\delta(x)$ with principle value $P,$ its real diagonal matrix elements
\begin{equation}
\label{shift}
\begin{split}
\Delta^P_{p}(\varepsilon_{p}^0)&\equiv\textrm{Re}[\widetilde\Sigma^{\bigstar}_R]^P_{pp}(\varepsilon_{p}^0)\\
&=\frac{e^2}{2m\Omega_{0}r^3}P\sum_{s}\langle p|-e{\bf x}|s\rangle\cdot{\bm\Lambda}\cdot\langle s|-e{\bf x}'|p\rangle\\
&\ \ \ \times\Big[\frac{\rho_s^0}{\varepsilon^0_s-\varepsilon_{p}^0-\hbar\Omega_{0}}+\frac{1-\rho_s^0}{\varepsilon^0_s-\varepsilon_{p}^0+\hbar\Omega_{0}}\Big]
%&\ \ \ \times\Big[\frac{1-\rho_s^0}{\varepsilon_p^0-\varepsilon^0_s-\hbar\Omega_{0}}+\frac{\rho_s^0}{\varepsilon_p^0-\varepsilon^0_s+\hbar\Omega_{0}}\Big]
\end{split}
\end{equation} 
account for the shifting of the $p$th orbital energy while its imaginary diagonal matrix elements 
\begin{equation}
\label{broad}
\begin{split}
-(&1/2)\Gamma^P_{pp}(\varepsilon_{p}^0)\equiv\textrm{Im}[\widetilde\Sigma^{\bigstar}_R]^P_{pp}(\varepsilon_{p}^0)\\
&=\frac{\pi e^2}{2m\Omega_{0}r^3}\sum_{s}\langle p|-e{\bf x}|s\rangle\cdot{\bm\Lambda}\cdot\langle s|-e{\bf x}'|p\rangle\\
&\ \ \ \times\big[(1-\rho_s^0)\delta(\varepsilon_s^0-\varepsilon^0_{p}+\hbar\Omega_{0})-\rho_s^0\delta(\varepsilon_s^0-\varepsilon^0_{p}-\hbar\Omega_{0})\big]
%&\ \ \ \ \ \ \ \times\left\{
%\begin{array}{c}
%-(1-\rho_s^0)\delta(\varepsilon_p^0-\varepsilon^0_s-\hbar\Omega_{0})\\
%+{\rho_s^0}\delta(\varepsilon_p^0-\varepsilon^0_s+\hbar\Omega_{0})
%\end{array}
%\right.
\end{split}
\end{equation} 
account for the rate of spontaneous emission of a plasmon with energy $\hbar\Omega_{0}$ from the electronically excited state $p$ (first term) or the rate of spontaneous absorption of a plasmon with energy $\hbar\Omega_{0}$ into the state $p$ (second term), both inducing molecular-electronic transitions to the state $s,$ where we have have assumed that $\Delta^P$ and $\Gamma^P$ vary so slowly with energy that we may choose $\hbar\omega=\varepsilon_{p}^0;$ both effects are due to the interaction between molecular electrons and the induced plasma density in the particle and are, here, rigorously included from first principles. From the point of view of the molecule, these interactions underlie a state-by-state broadening of the molecule's electronically excited states. It is precisely this interaction physics that is not explicitly treated in Ref. \cite{Jensen2006a,Aikens2006a,Jensen2007b}, but is, rather, implicitly encapsulated within a common phenomenological damping factor for all electronically-excited states. A generic consequence occurring whenever $\Gamma^P\neq0$ is that the effective Hamiltonian of the coupled molecule-particle system is no longer Hermitian. For completeness we note that both $\Delta^P$ and $\Gamma^P$ are related to each other by Hilbert transformation \cite{CT1992}.

%We note that, when the effects of damping due to the presence of the molecule and external field are included in the particle's polarizability, 
%It will be discussed in the following that, upon analytic continuation, the real poles of $\widetilde G_0$  $\widetilde {\cal G}^{P}$ has poles and 

This approximation of the metal particle's electrons as a homogeneous electron gas supporting collective excitation at the bulk plasma frequency is applied in the following to analytically demonstrate an enhanced Raman scattering from the coupled molecule-particle system in analogy to the classical result \cite{nitzan1980} reviewed in Sec. \ref{CSERS}.

\subsection{\label{TMATsec0}Scattering $T$-Matrix}
Transition amplitudes between initial and final eigenstates of an arbitrary reference Hamiltonian underlie the computation of many different observable quantities; such amplitudes are directly related to the scattering $T$-matrix which is the subject of this section. Recalling that time-dependent Green's functions are propagators in the sense that they describe the propagation of particles in time through an interacting many-particle assembly, it is not surprising that a connection exists between the exact one-body Green's function and the $S$-matrix of scattering theory; see, {\it e.g.} Ref. \cite{Newton1966a}. Specifically, their relationship for $t>0$ is given by
\begin{equation}
\label{Smat}
\begin{split}
{\cal S}_{fi}&=\lim_{t\to\infty}e^{i(\varepsilon_p+\varepsilon_q)t/2\hbar}{\tint}_{{\cal C}_+}\frac{dz}{2\pi i}\widetilde{\cal G}^R_{pq}(z)\\
&=\delta_{pq}-2\pi i\delta(\varepsilon_{p}-\varepsilon_{q}){\cal T}_{pq}
\end{split}
\end{equation} 
in the interaction representation, with one-body interaction $\hat{\cal V},$ where the effects of scattering from an initial many-body state with underlying one-body states labeled by $q$ to a final many-body state with underlying one-body states labeled by $p$ are encapsulated in the one-body $T$-matrix elements \cite{CT1992}
\begin{equation}
\label{genT}
{\cal T}_{fi}=\langle p|{\cal V}|q\rangle+\sum_{rs}\langle p|{\cal V}|r\rangle\widetilde{\cal G}^R_{rs}(\varepsilon_q)\langle s|{\cal V}|q\rangle.
\end{equation} 
Here the retarded Green's function $\widetilde{\cal G}^R(z)$ is analytically continued away from the real axis and into the complex $z$-plane where new features such as complex poles (also called resonances) and complex thresholds may be revealed on higher or lower Riemann sheets.

\begin{figure}[t]
%\psfrag{C}[][]{{\LARGE ${\cal C}_+$}}
%\psfrag{z1}[][]{{\LARGE $z_1$}}
%\psfrag{z2}[][]{{\LARGE $z_2$}}
%\psfrag{z3}[][]{{\LARGE $z_3$}}
\begin{center}
%\rotatebox{0}{\resizebox{!}{3.4cm}{\includegraphics{contour}}}
\rotatebox{0}{\resizebox{!}{3.4cm}{\includegraphics{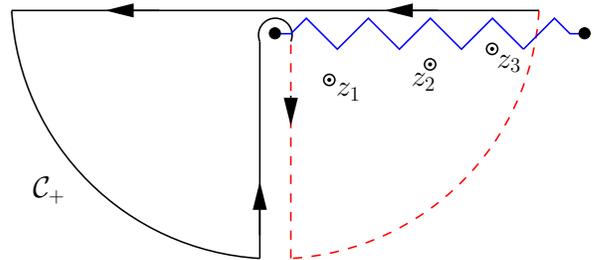}}}
\caption{\label{cont} (Color online) Complex contour associated with the $S$-matrix integral in Eq. (\ref{Smat}). The contour wraps around the branch point, where a molecular-electronic continuum channel opens due to interaction with the external field ${\bf E}_0,$ and passes onto the second Riemann sheet on the right-hand side; the associated part of ${\cal C}_+$ is indicated with a dashed red line. There it encloses complex poles (resonances) at the points $z_1,$ $z_2,$ and $z_3.$ A branch cut (blue) extends from the first threshold to $z=(+\infty,0).$}
\end{center}
\end{figure}

As we will be concerned with perturbations stemming from an external electric field ${\bf E}_0,$ we make the dipole-interaction approximation and take $\hat{\cal V}=-\hat{\bf d}\cdot{\bf E}_0$ for the purposes of the present discussion. The contour ${\cal C}_+,$ which is displayed in Fig. \ref{cont}, is rerouted to avoid the branch point resulting from the (real) threshold where an electronic continuum channel opens due to the absorption/emission of a photon from/to the field ${\bf E}_0.$ An associated branch cut connects this branch point to its terminal branch point chosen at $z=(+\infty,0).$ As a result of this particular route for ${\cal C}_+,$ the contour integration moves onto the second Riemann sheet on the right-hand side (shown in red in Fig. \ref{cont}), where it encloses resonances at the points $z_1,$ $z_2,$ and $z_3$ (assuming, for the purpose of demonstration, that only three exist). What were real eigenenergies of the unperturbed Hamiltonian now become complex eigenenergies ($\textrm{Im}z_j\neq0,$ $j=1,2,3$) of the interacting system. The particular locations of these complex poles of the exact one-body Green's function are due to the analytic continuation of $\widetilde{\cal G}^R$ from the upper-half plane onto the second Riemann sheet in lower-half plane.
%!!!!!!!!!!!!!!!!!!!!!!!!!!!!!!!!!!!!!!!!!!!!!!!!!!!!!!!!!!!
%!!!!!!!!!!!!!!!!!!!!!!!!!!!!!!!!!!!!!!!!!!!!!!!!!!!!!!!!!!!
%!!!!!!!!!!!!!!!!!!!!!!!!!!!!!!!!!!!!!!!!!!!!!!!!!!!!!!!!!!!
%!!!!!!!!!!!!!!!!!!!!!!!!!!!!!!!!!!!!!!!!!!!!!!!!!!!!!!!!!!!

\subsubsection{\label{TMATsec1}Normal Raman scattering from a noninteracting molecular system}
We are now in a position to compute Raman transition amplitudes between states of the interacting molecule-particle system induced by the perturbation $\hat{\cal V}=\hat F_E=-\hat{\bf d}\cdot[{\bf E}_0+{\bf E}_P].$ However, before computing the associated interacting scattering $T$-matrix, we first make a detour and consider the case of normal Raman scattering from an isolated Raman-active molecule using the many-body Green's function formalism. A more thorough theoretical development of Raman scattering that does not involve Green's functions can be found in Ref. \cite{Craig}, while an advanced review covering linear and nonlinear optical processes from a Green's function perspective can be found in Ref. \cite{Mukamel1995}. Here, noninteracting molecular electrons are described at the level of HF mean-field theory by the one-body HF Green's function $G^{(0)}$ defined previously in Eqs. (\ref{Gtime}) and (\ref{LG}); molecular-nuclear degrees of freedom $J=1,\ldots,M,$ which underlie all electronic states within the Born-Oppenheimer approximation, are represented by the one-body vibrational states $|\nu_{J}\rangle.$ In the harmonic approximation, the total $M$-body molecular-vibrational wave function is equal to the unsymmetrized (Hartree) product $\prod_{J=1}^M\langle{\bf Q}_J|\nu_{J}\rangle$ of one-body vibrational wave functions $\langle{\bf Q}_J|\nu_{J}\rangle$ for each degree of freedom. Molecular rotations are not resolved in our presentation. Perturbed by the external field ${\bf E}_{0},$ the scattering $T$-matrix of the molecular system may be approximated at second order in the dipole-interaction perturbation $\hat{\cal V}=-\hat{\bf d}\cdot{\bf E}_{0}$ as
\begin{widetext}
\begin{equation}
\label{HFTapp}
\begin{split}
\langle\nu^{\prime}_{J}|&{\cal T}^{\textrm{HF}(2)}_{fi}|\nu_{J}\rangle\\
&=\sum_{rs}\langle\nu^{\prime}_{J}|\Big[\widetilde{\bf E}_{0}^{(-)}\cdot\langle p|e{\bf x}|r\rangle[\widetilde{G}^R_{(0)}]_{rs}(\varepsilon_{q}^0+\hbar\omega_{\bf k})\langle s|e{\bf x}'|q\rangle\cdot\widetilde{\bf E}_{0}^{(+)}+\widetilde{\bf E}_{0}^{(+)}\cdot\langle p|e{\bf x}|r\rangle[\widetilde{G}^R_{(0)}]_{rs}(\varepsilon_{q}^0-\hbar\omega_{{\bf k}'})\langle s|e{\bf x}'|q\rangle\cdot\widetilde{\bf E}_{0}^{(-)}\Big]|\nu_{J}\rangle\\
&\approx\sum_{rs}\widetilde{E}_{0\xi}^{(-)}\langle\nu^{\prime}_{J}|\Big[\frac{\langle p|e{x}^\xi|r\rangle\delta_{rs}\langle s|e{x}^{\prime\sigma}|q\rangle}{\varepsilon_{q}^0-\varepsilon_{s}^0+\hbar\omega_{\bf k}+i0^+}+\frac{\langle p|e{x}^{\sigma}|r\rangle\delta_{rs}\langle s|e{x}^{\prime\xi}|q\rangle}{\varepsilon_{q}^0-\varepsilon_{s}^0-\hbar\omega_{{\bf k}'}+i0^+}\Big]|\nu_{J}\rangle\widetilde{E}_{0\sigma}^{(+)},
%&=\sum_{s}\Big[\widetilde{\bf E}_{0}^{(-)}\cdot\frac{\langle p|e{\bf x}|s\rangle\langle s|e{\bf x}'|q\rangle}{\varepsilon_q^0-\varepsilon_s^0+\hbar\omega_{\bf k}+i0^+}\cdot\widetilde{\bf E}_{0}^{(+)}\\
%&\hspace{1cm}+\widetilde{\bf E}_{0}^{(+)}\cdot\frac{\langle p|e{\bf x}|s\rangle\langle s|e{\bf x}'|q\rangle}{\varepsilon_q^0-\varepsilon_s^0-\hbar\omega_{{\bf k}'}+i0^+}\cdot\widetilde{\bf E}_{0}^{(-)}\Big],
%&=\widetilde{\bf E}_{0}^{(-)}\cdot\widetilde{\bm\alpha}_{pq}(\hbar\omega_q)\cdot\widetilde{\bf E}_{0}^{(+)}
\end{split}
\end{equation}
\end{widetext}
which is expressed in terms of the retarded HF Green's function $\widetilde G^R_{(0)}$ (\ref{retG}). Here, in addition to the sum over the intermediate electronic states $r$ and $s,$ there should be a sum over the intermediate vibrational states of the molecule; however, for simplicity in presentation, we omit the vibrational energy differences in the denominator (in comparison to the electronic energy differences) here and in the following and appeal to the closure relation $1=\sum_\nu|\nu_J\rangle\langle\nu_J|$ in the numerator \cite{Craig}. It is also important to note that, due to Kronecker delta $\delta_{rs}$ in the numerator, the intermediate electronic states must be the same and, further, must label either particle-particle or hole-hole states; no particle-hole or hole-particle intermediate states contribute to Raman scattering.  This point will be important in deriving an expression for the SERS intensity in Sec. \ref{TMATsec2} below.

The one-body states which underlie the initial and final molecular states for normal Raman scattering are
\begin{equation}
\label{if}
\begin{split}
|q\rangle|\nu_{J}\rangle&\equiv|\phi_q;\{\nu_{J};{\cal N}_{{\bf k}\lambda},{\cal N}'_{{\bf k}'\lambda'}\}\rangle|\nu_{J}\rangle\\
|p\rangle|\nu^{\prime}_{J}\rangle&\equiv|\phi_p;\{\nu^{\prime}_{J};({\cal N}-1)_{{\bf k}\lambda},({\cal N}'+1)_{{\bf k}'\lambda'}\}\rangle|\nu^{\prime}_{J}\rangle,
\end{split}
\end{equation}
where the incident and Raman-scattered fields are implicitly labeled in the molecular-electronic states only to motivate proper field normalization; here ${\cal N}$ (${\cal N}-1$) photons are initially (finally) in the state $|{\bf k};\lambda\rangle$ with wave vector ${\bf k},$ polarization $\lambda,$ and energy $\hbar\omega_{\bf k},$ and ${\cal N}'$ (${\cal N}'+1$) photons are initially (finally) in the state $|{\bf k}';\lambda'\rangle$ with wave vector ${\bf k}',$ polarization $\lambda',$ and energy $\hbar\omega_{{\bf k}'}.$ As previously discussed, within the Born-Oppenheimer approximation, the molecular-electronic and nuclear coordinates are separated as $|q;\{\nu_{J}\}\rangle|\nu_{J}\rangle$ and $|p;\{\nu^{\prime}_{J}\}\rangle|\nu^{\prime}_{J}\rangle,$ where $\nu_{J}$ and $\nu^{\prime}_{J}$ label the initial and final vibrational quanta associated with the particular normal-mode coordinate ${\bf Q}_J.$

\begin{figure}[t]
%\psfrag{p}[][]{{\Huge $p$}}
%\psfrag{q}[][]{{\Huge $q$}}
%\psfrag{a}[][]{{\Huge $\widetilde\alpha^M_{fi}$}}
%\psfrag{v1}[][]{{\Huge $\nu_{J}$}}
%\psfrag{v2}[][]{{\Huge $\nu_{J}^\prime$}}
%\psfrag{E01}[][]{{\Huge $\widetilde{\bf E}_0^{(+)}$}}
%\psfrag{E02}[][]{{\Huge $\widetilde{\bf E}_0^{(-)}$}}
%\psfrag{G}[][]{{\Huge $\widetilde{\cal G}^{\textrm{int}(2)}_{pq}(\omega)-\widetilde{\cal G}^P_{pq}(\omega)$}}
\begin{center}
%\rotatebox{0}{\resizebox{!}{4.1cm}{\includegraphics{vib}}}
\rotatebox{0}{\resizebox{!}{4.1cm}{\includegraphics{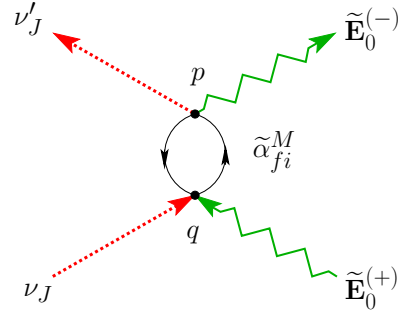}}}
\caption{\label{vib0} (Color online) Diagrammatic representation of the processes occurring in normal Raman scattering from a molecular target with electronic transition polarizability $\widetilde\alpha^M_{fi}.$ Molecular vibrational states (red) are labeled by $\nu_{J}$ and $\nu'_{J}$ and the incident and Raman-scattered electric field (green) by $\widetilde{\bf E}_0^{(+)}$ and $\widetilde{\bf E}_0^{(-)}.$ Both are implicitly accounted for in our semiclassical approach through the state labels.}
\end{center}
\end{figure}

Recognizing the inverse of the retarded HF polarization propagator $\widetilde\Pi^R_{(0)}$ [Eq. (\ref{retPi})] in the denominator of Eq. (\ref{HFTapp}) and recalling the connection between $\widetilde\Pi_{(0)}^R$ and the linear polarizability defined in Eq. (\ref{transalpha}), we find that 
\begin{equation}
\label{HFT2}
\begin{split}
\langle&\nu^{\prime}_{J}|{\cal T}^{\textrm{HF}(2)}_{fi}|\nu_{J}\rangle\\
&=-\widetilde{\bf E}_{0}^{(-)}\cdot\langle\nu^{\prime}_{J}|\widetilde{\bm\alpha}^M_{fi}(\hbar\omega_{\bf k},-\hbar\omega_{{\bf k}'})|\nu_{J}\rangle\cdot\widetilde{\bf E}_{0}^{(+)}\\
&=-(-i)\sqrt{\frac{2\pi\hbar\omega_{{\bf k}'}({\cal N}'+1)_{{\bf k}'\lambda'}}{L^3}}i\sqrt{\frac{2\pi\hbar\omega_{\bf k}{\cal N}_{{\bf k}\lambda}}{L^3}}\\
&\hspace{1.2cm}\times\hat{\bm\epsilon}^{(-)}_{\lambda'}({\bf k}')\cdot\langle\nu^{\prime}_{J}|\widetilde{\bm\alpha}_{fi}^M(\hbar\omega_{\bf k},-\hbar\omega_{{\bf k}'})|\nu_{J}\rangle\cdot\hat{\bm\epsilon}^{(+)}_\lambda({\bf k}),
\end{split}
\end{equation}
where, as in Sec. \ref{CSERS}, the incident electric field amplitude is $i\sqrt{2\pi\hbar\omega_{\bf k}{\cal N}_{{\bf k}\lambda}/{L^3}}\hat{\bm\epsilon}^{(+)}_\lambda({\bf k})$ as is consistent with the field occupation numbers in Eq. (\ref{if}). For completeness we point out that had ${\bf E}_{0}$ been properly treated as a quantum-mechanical field, the number of scattered photons in Eq. (\ref{HFT2}) would have rigorously been $({\cal N}'+1)_{{\bf k}'\lambda'};$ see, {\it e.g.}, Ref. \cite{Ratner}. A diagrammatic representation of the electric field interaction and nuclear vibrational processes occurring in normal Raman scattering is displayed in Fig. \ref{vib0}.
%!!!!!!!!!!!!!!!!!!!!!!!!!!!!!!!!!!!!!!!!!!!!!!!!!!!!!!!!!!!
%!!!!!!!!!!!!!!!!!!!!!!!!!!!!!!!!!!!!!!!!!!!!!!!!!!!!!!!!!!!
%!!!!!!!!!!!!!!!!!!!!!!!!!!!!!!!!!!!!!!!!!!!!!!!!!!!!!!!!!!!
%!!!!!!!!!!!!!!!!!!!!!!!!!!!!!!!!!!!!!!!!!!!!!!!!!!!!!!!!!!!

Fermi's golden rule of time-dependent perturbation theory \cite{Merz} dictates that the rate of transition $w$ between states $f$ and $i$ is related to the transition amplitude by 
\begin{equation}
w_{fi}^{\textrm{HF}(2)}=\frac{2\pi}{\hbar}\varrho(\hbar\omega_{{\bf k}'})\big|\langle\nu^{\prime}_{J}|{\cal T}^{\textrm{HF}(2)}_{fi}|\nu_{J}\rangle\big|^2
\end{equation} 
for the particular case of normal Raman scattering from a noninteracting molecular system, where $\varrho(\hbar\omega_{{\bf k}'})=(L/2\pi c)^3(\omega_{{\bf k}'}^2/\hbar)d\Omega_{{\bf k}'}$ is the density of states of the emitted electric field propagating in the ${\bf k}'$-direction. It is now straightforward to compute the Raman-scattering intensity in the direction ${\bf k}'$ from a single molecular scatterer to be
\begin{equation}
\begin{split}
&\frac{I_{\textrm{Raman}}({\bf k}')}{I_0({\bf k})}=\frac{\omega_{\bf k}\omega^3_{{\bf k}'}}{c^4}({\cal N}'+1)_{{\bf k}'\lambda'}\Big|\sum_J\hat{\bm\epsilon}^{(-)}_{\lambda'}({\bf k}')\\
&\ \cdot\langle\nu^{\prime}_{J}|({\bf Q}_J-{\bf Q}_{0})\cdot\nabla_{{\bf Q}_J}\widetilde{\bm\alpha}^M_{fi}(\hbar\omega_{\bf k},-\hbar\omega_{{\bf k}'})|\nu_{J}\rangle\cdot\hat{\bm\epsilon}^{(+)}_{\lambda}({\bf k})\Big|^2,
\end{split}
\end{equation} 
where $I_0({\bf k})$ is the intensity of the incident field in the direction ${\bf k}$ with polarization $\lambda,$ and where only the linear term in the Taylor expansion of the electronic polarizability 
\begin{equation}
\widetilde{\bm\alpha}_M[\{{\bf Q}\}]=\widetilde{\bm\alpha}_M[{\bf Q}_0]+\sum_J({\bf Q}_J-{\bf Q}_{0})\cdot\nabla_{{\bf Q}_J}\widetilde{\bm\alpha}_M[{\bf Q}_0]+\cdots
\end{equation} 
around the molecular equilibrium geometry ${\bf Q}_0$ was retained. Here the sum runs over all $M$ normal-mode coordinates $\{{\bf Q}\}$ of the molecule and it is simple to show that $\langle\nu^{\prime}_{J}|({\bf Q}_J-{\bf Q}_{0})|\nu_{J}\rangle$ is nonzero whenever $\nu^{\prime}_{J}=\nu_{J}\pm1.$ Elastic Rayleigh scattering is described by the first term, while inelastic Raman scattering, at lowest order, occurs through the second. Raman-scattering overtones beyond the fundamental depend upon higher-order terms. This expression is the quantum-mechanical analog of the classical expression derived in Eq. (\ref{I1}) for a polarizable molecule interacting with the external field ${\bf E}_{0}.$

\subsubsection{\label{TMATsec2}Enhanced Raman scattering from an interacting system}
Having briefly reviewed the theory of normal Raman scattering from a noninteracting molecular-electronic system within the many-body Green's function formalism, we now turn to the case where the molecule is itself interacting with a nearby metal particle, including both image and local-field effects. The one-body Green's function ${\cal G}^P$ developed in Sec. \ref{mep} was designed specifically to incorporate this physics; it includes the self-induced polarization effects of a nearby classical metallic particle to infinite order in perturbation theory ({\it i.e.}, the image effects) and, for the purpose of demonstration only, assumes that the conduction electrons of the metal are well-described by the RPA. Comparison of the expression (\ref{genT}) for the scattering $T$-matrix with the second-order perturbative approximation to the irreducible self energy $\hbar\widetilde\Sigma_{E}^{\bigstar}$ displayed in Eq. (\ref{ISEE}) shows that
\begin{widetext}
\begin{equation}
\label{intTmat}
\begin{split}
&\langle\nu^{\prime}_{J}|{\cal T}^{\textrm{int}(2)}_{fi}|\nu_{J}\rangle=\langle\nu^{\prime}_{J}|\hbar[\widetilde\Sigma_{R}^{\bigstar}]_E(\varepsilon_{q}^0+\hbar\omega_{\bf k})|\nu_{J}\rangle+\langle\nu^{\prime}_{J}|\hbar[\widetilde\Sigma_{R}^{\bigstar}]_E(\varepsilon_{q}^0-\hbar\omega_{{\bf k}'})|\nu_{J}\rangle\\
&=\sum_{rs;XY}\langle\nu^{\prime}_{J}|\Big[\widetilde{\bf E}_X^{(-)}\cdot\langle p|e{\bf x}|r\rangle[\widetilde{\cal G}^R_{P}]_{rs}(\varepsilon_{q}^0+\hbar\omega_{\bf k})\langle s|e{\bf x}'|q\rangle\cdot\widetilde{\bf E}_Y^{(+)}+\widetilde{\bf E}_Y^{(+)}\cdot\langle p|e{\bf x}|r\rangle[\widetilde{\cal G}^R_{P}]_{rs}(\varepsilon_{q}^0-\hbar\omega_{{\bf k}'})\langle s|e{\bf x}'|q\rangle\cdot\widetilde{\bf E}_X^{(-)}\Big]|\nu_{J}\rangle,
\end{split}
\end{equation}
\end{widetext}
which is defined in terms of the retarded interacting one-body Green's function $\widetilde{\cal G}^R_{P},$ and where the energy of the incident field is $\hbar\omega_{\bf k}$ \cite{assume}. The underlying one-body states associated with enhanced Raman scattering are similar to those previously defined in Eq. (\ref{if}) for normal Raman scattering, {\it i.e.,}
\begin{equation}
\label{ifenh}
\begin{split}
|q\rangle|\nu_{J}\rangle&\equiv|\chi_q;\{\nu_{J};{\cal N}_{{\bf k}\lambda},{\cal N}'_{{\bf k}'\lambda'}\}\rangle|\nu_{J}\rangle\\
|p\rangle|\nu^{\prime}_{J}\rangle&\equiv|\chi_p;\{\nu^{\prime}_{J};({\cal N}-1)_{{\bf k}\lambda},({\cal N}'+1)_{{\bf k}'\lambda'}\}\rangle|\nu^{\prime}_{J}\rangle.
\end{split}
\end{equation}
As previous, they label both electronic and nuclear vibrational states of the molecule. However, here, the one-body electronic states are described by the interacting Dyson orbitals $\chi_q,$ which are solutions of the Dyson equation (\ref{phiint1}), rather than by the noninteracting HF orbitals $\phi_q.$ The labels $X,Y\in\{0,0P\}$ refer to either the external field $\widetilde{\bf E}_{0}^{(\pm)}$ or to the field of the particle $\widetilde{\bf E}_{0P}^{(\pm)}.$ From the two terms in Eq. (\ref{intTmat}) there are eight possible ways to arrange these two fields between the two states $X,Y:$ four from the first term (uncrossed interactions) and four from the second term (crossed interactions). All eight terms are included in the formalism; they enumerate all possible time orderings between uncrossed and crossed interactions. In this sense, the scattering theory description is noncausal with each interaction event equally as important as all others \cite{Bialynicki-Birula2007,Mukamel2007}. Note that this expression constitutes the Born approximation to the field perturbation as $\widetilde{\cal G}^{P}$ itself does not include any effects of field interaction. Dyson's expansion of $\widetilde{\cal G}^{P}$ subjected to the field would provide a systematic way to build in these effects perturbatively.

From Eq. (\ref{intTmat}), the Born approximation to the interacting scattering $T$-matrix with respect to the external fields ${\bf E}_{0}$ and ${\bf E}_P$ is given by
\begin{widetext}
\begin{equation}
\label{TSERS0}
\begin{split}
\langle\nu^{\prime}_{J}|{\cal T}^{\textrm{int}(2)}_{fi}|\nu_{J}\rangle&=\sum_{rs=ij,ab;XY}\langle\nu^{\prime}_{J}|\Big\{\widetilde{\bf E}_X^{(-)}\cdot\langle p|e{\bf x}|r\rangle\frac{\delta_{rs}}{\hbar\omega_{\bf k}+\varepsilon_{q}^0-\varepsilon_{r}^0-\hbar[\widetilde{\Sigma}^{\bigstar}_R]^P_{rr}(\varepsilon_{q}^0+\hbar\omega_{\bf k})}\langle s|e{\bf x}'|q\rangle\cdot\widetilde{\bf E}_Y^{(+)}\\
&\ \ \ +\widetilde{\bf E}_X^{(+)}\cdot\langle p|e{\bf x}|r\rangle\frac{\delta_{rs}}{-\hbar\omega_{{\bf k}'}+\varepsilon_{q}^0-\varepsilon_{r}^0-\hbar[\widetilde{\Sigma}_R^{\bigstar}]_{rr}^P(\varepsilon_{q}^0-\hbar\omega_{{\bf k}'})}\langle s|e{\bf x}'|q\rangle\cdot\widetilde{\bf E}_Y^{(-)}\Big\}|\nu_{J}\rangle\\
&+\sum_{rs=ib,aj;XY}\langle\nu^{\prime}_{J}|\Big\{\widetilde{\bf E}_X^{(-)}\cdot\langle p|e{\bf x}|r\rangle[\widetilde{\cal G}^R_{P}]_{rs}(\varepsilon_{q}^0+\hbar\omega_{\bf k})\langle s|e{\bf x}'|q\rangle\cdot\widetilde{\bf E}_Y^{(+)}\\
&\ \ \ +\widetilde{\bf E}_Y^{(+)}\cdot\langle p|e{\bf x}|r\rangle[\widetilde{\cal G}^R_{P}]_{rs}(\varepsilon_{q}^0-\hbar\omega_{{\bf k}'})\langle s|e{\bf x}'|q\rangle\cdot\widetilde{\bf E}_X^{(-)}\Big\}|\nu_{J}\rangle,\\
\end{split}
\end{equation}
\end{widetext}
where the off-diagonal matrix elements of the interacting molecular Green's function $\widetilde{\cal G}^R_{P}$ in the last two terms were given in Eq. (\ref{G2M}). As discussed previously below Eq. (\ref{HFTapp}), the intermediate states associated with Raman scattering are restricted to particle-particle and hole-hole states; no particle-hole or hole-particle intermediate states contribute to its lowest-order theoretical description. Consequently, in the following, we omit the off-diagonal components of $\widetilde{\cal G}^R_{P}$ underlying the interacting scattering $T$-matrix. The first (diagonal) term stemming from $\hbar\widetilde\Sigma_{E}^{\bigstar}(\varepsilon_{q}^0+\hbar\omega_{\bf k})$ represents the uncrossed scattering contributions analogous to those shown in Fig. \ref{ram0}, while the second (diagonal) term stemming from $\hbar\Sigma_{E}^{\bigstar}(\varepsilon_{q}^0-\hbar\omega_{{\bf k}'})$ represents the crossed terms (not shown) where a Raman-scattered photon is emitted in the initial molecular state and an incident photon strikes in the final state.

By including only particle-particle and hole-hole intermediate molecular-electronic states, the interacting scattering $T$-matrix reduces to
\begin{widetext}
\begin{equation}
\label{TSERS}
\begin{split}
\langle\nu^{\prime}_{J}|&{\cal T}^{\textrm{int}(2)}_{fi}|\nu_{J}\rangle\\
&\approx\sum_{rs=ij,ab;XY}\langle\nu^{\prime}_{J}|\widetilde{E}_{X\xi}^{(-)}\Bigl[\frac{\langle p|e{x}^\xi|r\rangle\delta_{rs}\langle s|e{x}^{\prime\sigma}|q\rangle}{\hbar\omega_{\bf k}+\varepsilon_{q}^0-\varepsilon_r^0-\hbar[\widetilde{\Sigma}^{\bigstar}_R]_{rr}^P(\varepsilon_{q}^0+\hbar\omega_{\bf k})}+\frac{\langle p|e{x}^\sigma|r\rangle\delta_{rs}\langle s|e{x}^{\prime\xi}|q\rangle}{-\hbar\omega_{{\bf k}'}+\varepsilon_{q}^0-\varepsilon_r^0-\hbar[\widetilde{\Sigma}^{\bigstar}_R]_{rr}^P(\varepsilon_{q}^0-\hbar\omega_{{\bf k}'})}\Bigr]\widetilde{E}_{Y\sigma}^{(+)}|\nu_{J}\rangle\\
&=\sum_{rs=ij,ab;XY}\langle\nu^{\prime}_{J}|\widetilde{E}_{X\xi}^{(-)}\Bigl[\frac{\langle p|e{x}^\xi|r\rangle\delta_{rs}[\widetilde\Pi_{(0)}^{R}]_{rq}(\hbar\omega_{\bf k})\langle s|e{x}^{\prime\sigma}|q\rangle}{1-[\widetilde\Pi_{(0)}^{R}]_{rq}(\hbar\omega_{\bf k})\hbar[\widetilde{\Sigma}^{\bigstar}_R]^P_{rr}(\varepsilon_{q}^0+\hbar\omega_{\bf k})}+\frac{\langle p|e{x}^\sigma|r\rangle\delta_{rs}[\widetilde\Pi_{(0)}^{R}]_{rq}(-\hbar\omega_{{\bf k}'})\langle s|e{x}^{\prime\xi}|q\rangle}{1-[\widetilde\Pi_{(0)}^{R}]_{rq}(-\hbar\omega_{{\bf k}'})\hbar[\widetilde{\Sigma}^{\bigstar}_R]^P_{rr}(\varepsilon_{q}^0-\hbar\omega_{{\bf k}'})}\Bigr]\widetilde{E}_{Y\sigma}^{(+)}|\nu_{J}\rangle\\
%&\approx\sum_{rs=fi;XY}\widetilde{E}_{X\xi}^{(-)}\frac{\langle f|e{x}^\xi|r\rangle\delta_{rs}[\widetilde\Pi_{(0)}^{R}]_{ri}(\hbar\omega_{\bf k})\langle s|e{x}^{\prime\sigma}|i\rangle+\langle f|e{x}^\sigma|r\rangle\delta_{rs}[\widetilde\Pi_{(0)}^{R}]_{ri}(-\hbar\omega_{{\bf k}'})\langle s|e{x}^{\prime\xi}|i\rangle}{\{1-[\widetilde\Pi_{(0)}^{R}]_{ri}(\hbar\omega_{\bf k})\hbar\widetilde{\Sigma}_{rr}^{\bigstar P}(\varepsilon_{i}^0+\hbar\omega_{\bf k})\}\{1-[\widetilde\Pi_{(0)}^{R}]_{ri}(-\hbar\omega_{{\bf k}'})\hbar\widetilde{\Sigma}_{rr}^{\bigstar P}(\varepsilon_{i}^0-\hbar\omega_{{\bf k}'})\}}\widetilde{E}_{Y\sigma}^{(+)}\\
&\approx\sum_{rs=ij,ab;XY}\widetilde{E}_{X\xi}^{(-)}\langle\nu^{\prime}_{J}|\frac{-[\widetilde\alpha_M]_{pq,rs}^{\xi\sigma}(\hbar\omega_{\bf k},-\hbar\omega_{{\bf k}'})\delta_{rs}}{\{1-[\widetilde\Pi_{(0)}^{R}]_{rq}(\hbar\omega_{\bf k})\hbar[\widetilde{\Sigma}^{\bigstar}_R]_{rr}^P(\varepsilon_{q}^0+\hbar\omega_{\bf k})\}\{1-[\widetilde\Pi_{(0)}^{R}]_{rq}(-\hbar\omega_{{\bf k}'})\hbar[\widetilde{\Sigma}^{\bigstar}_R]^P_{rr}(\varepsilon_{q}^0-\hbar\omega_{{\bf k}'})\}}|\nu_{J}\rangle\widetilde{E}_{Y\sigma}^{(+)},
\end{split}
\end{equation}
\end{widetext}
where two terms in the numerator which stem from finding a common denominator have been omitted in the third line. As with the off-diagonal terms, these terms involve higher powers of the molecular polarizability and correspond to repeated photon scattering events with the molecule. Identification of the inverse retarded HF polarization propagator from Eq. (\ref{retPi}) in the denominator of this expression has been made and can be used to simplify the denominator in the last line as
\begin{equation}
\label{den}
\begin{split}
[\widetilde\Pi_{(0)}^{R}]&_{rq}(\pm\hbar\omega_{{\bf k},{\bf k}'})\hbar[\widetilde{\Sigma}^{\bigstar}_R]_{rr}^P(\varepsilon_{q}^0\pm\hbar\omega_{{\bf k},{\bf k}'})\\
&=\sum_{tu}{\textrm{Tr}}\big\{\widetilde{\bm\alpha}^{M}_{rq,ut}(\pm\hbar\omega_{{\bf k},{\bf k}'})\cdot{\bm\Lambda}\\
&\hspace{0.5cm}\cdot{\tint}\frac{d\varepsilon'}{2\pi i}\hbar^2\widetilde{\bm\alpha}_{\textrm{RPA}}(\varepsilon')[\widetilde G^R_{(0)}]_{tu}(\varepsilon_{q}^0\pm\hbar\omega_{{\bf k},{\bf k}'}+\varepsilon')\cdot{\bm\Lambda}\big\}.
\end{split}
\end{equation} 
Within the RPA, the integral in this expression is the retarded component of the integral computed previously in Eq. (\ref{RPAISE}). Two different transition polarizabilities appear in Eqs. (\ref{TSERS}) and (\ref{den}); in light of Eq. (\ref{transalpha}) they are
\begin{equation}
\begin{split}
-&\widetilde{\bm\alpha}^M_{pq,rs}(\hbar\omega_{\bf k},-\hbar\omega_{{\bf k}'})=\langle p|e{\bf x}|r\rangle[\widetilde\Pi_{(0)}^{R}]_{rq}(\hbar\omega_{\bf k})\langle s|e{\bf x}'|q\rangle\\
&\hspace{2.75cm}+\langle p|e{\bf x}|r\rangle[\widetilde\Pi_{(0)}^{R}]_{rq}(-\hbar\omega_{{\bf k}'})\langle s|e{\bf x}'|q\rangle\\
&\widetilde{\bm\alpha}^M_{rq,ut}(\pm\hbar\omega_{{\bf k},{\bf k}'})=\langle u|e{\bf x}|r\rangle[\widetilde\Pi_{(0)}^{R}]_{rq}(\pm\hbar\omega_{{\bf k},{\bf k}'})\langle r|e{\bf x}'|t\rangle.
\end{split}
\end{equation}

Performing the summation over $X,Y$ results in the following expression for the interacting scattering $T$-matrix
\begin{widetext}
\begin{equation}
\label{TSERS2}
\begin{split}
\langle\nu^{\prime}_{J}|{\cal T}^{\textrm{int}(2)}_{fi}|\nu_{J}\rangle&=-\sum_{rs=ij,ab}\langle\nu^{\prime}_{J}|\frac{\delta_{rs}}{\big\{1-[\widetilde\Pi_{(0)}^{R}]_{rq}(\hbar\omega_{\bf k})\hbar[\widetilde{\Sigma}^{\bigstar}_R]^P_{rr}(\varepsilon_{q}^0+\hbar\omega_{{\bf k}})\big\}\big\{1-[\widetilde\Pi_{(0)}^{R}]_{rq}(-\hbar\omega_{{\bf k}'})\hbar[\widetilde{\Sigma}^{\bigstar}_R]^P_{rr}(\varepsilon_{q}^0-\hbar\omega_{{\bf k}'})\big\}}\\
&\hspace{1cm}\times\Big[\widetilde{\bf E}_{0}^{(-)}\cdot\widetilde{\bm\alpha}^M_{pq,rs}(\hbar\omega_{\bf k},-\hbar\omega_{{\bf k}'})\cdot\widetilde{\bf E}_{0}^{(+)}+\widetilde{\bf E}_{0P}^{(-)}(-\omega_{{\bf k}'})\cdot\widetilde{\bm\alpha}^M_{pq,rs}(\hbar\omega_{\bf k},-\hbar\omega_{{\bf k}'})\cdot\widetilde{\bf E}_{0}^{(+)}\\
&\hspace{1.5cm}+\widetilde{\bf E}_{0}^{(-)}\cdot\widetilde{\bm\alpha}^M_{pq,rs}(\hbar\omega_{\bf k},-\hbar\omega_{{\bf k}'})\cdot\widetilde{\bf E}_{0P}^{(+)}(\omega_{\bf k})+\widetilde{\bf E}_{0P}^{(-)}(-\omega_{{\bf k}'})\cdot\widetilde{\bm\alpha}^M_{pq,rs}(\hbar\omega_{\bf k},-\hbar\omega_{{\bf k}'})\cdot\widetilde{\bf E}_{0P}^{(+)}(\omega_{\bf k})\Big]|\nu_{J}\rangle\\
%&=\sum_{rstu}\frac{\delta_{rs}+f_{q,usrt}(\widetilde{\bm\alpha}_M,\widetilde{\bm\alpha}_{\textrm{RPA}})}{\big\{1-[\widetilde\Pi_{(0)}^{R}]_{rq}(\hbar\omega_{\bf k})\hbar\widetilde{\Sigma}_{rr}^{\bigstar P}\big\}\big\{1-[\widetilde\Pi_{(0)}^{R}]_{sq}(\hbar\omega_{\bf k})\hbar\widetilde{\Sigma}_{ss}^{\bigstar P}\big\}}\Big[\widetilde{\bf E}_{0}^{(-)}\cdot\widetilde{\bm\alpha}^M_{pq,rs}(\omega_{\bf k})\cdot\widetilde{\bf E}_{0}^{(+)}+\widetilde{\bf E}^{(-)}_0\cdot\widetilde{\bm\alpha}_{\textrm{RPA}}(-\omega_{{\bf k}'})\cdot{\bm\Lambda}\cdot\widetilde{\bm\alpha}^M_{pq,rs}(\omega_{\bf k})\cdot\widetilde{\bf E}_{0}^{(+)}\\
%&\hspace{2.5cm}+\widetilde{\bf E}_{0}^{(-)}\cdot\widetilde{\bm\alpha}^M_{pq,rs}(\omega_{\bf k})\cdot{\bm\Lambda}\cdot\widetilde{\bm\alpha}_{\textrm{RPA}}(\omega_{\bf k})\cdot\widetilde{\bf E}^{(+)}_0+\widetilde{\bf E}^{(-)}_0\cdot\widetilde{\bm\alpha}_{\textrm{RPA}}(-\omega_{{\bf k}'})\cdot{\bm\Lambda}\cdot\widetilde{\bm\alpha}^M_{pq,rs}(\omega_{\bf k})\cdot{\bm\Lambda}\cdot\widetilde{\bm\alpha}_{\textrm{RPA}}(\omega_{\bf k})\cdot\widetilde{\bf E}^{(+)}_0\Big]\\
&=-\sum_{r}\widetilde{E}_{0\xi}^{(-)}\langle\nu^{\prime}_{J}|[\widetilde{\alpha}_M]^{\gamma\eta}_{pq,rr}(\hbar\omega_{\bf k},-\hbar\omega_{{\bf k}'})|\nu_{J}\rangle\widetilde{E}_{0\sigma}^{(+)}\\
&\hspace{1cm}\times\frac{\delta^{\xi\gamma}\delta^{\eta\sigma}+\widetilde{\alpha}_{\textrm{RPA}}^{\xi\delta}(-\omega_{{\bf k}'}){\Lambda}^{\delta\gamma}\delta^{\eta\sigma}+\delta^{\xi\gamma}{\Lambda}^{\eta\beta}\widetilde{\alpha}_{\textrm{RPA}}^{\beta\sigma}(\omega_{\bf k})+\widetilde{\alpha}_{\textrm{RPA}}^{\xi\delta}(-\omega_{{\bf k}'}){\Lambda}^{\delta\gamma}{\Lambda}^{\eta\beta}\widetilde{\alpha}_{\textrm{RPA}}^{\beta\sigma}(\omega_{\bf k})}{\big\{1-[\widetilde\Pi_{(0)}^{R}]_{rq}(\hbar\omega_{\bf k})\hbar[\widetilde{\Sigma}^{\bigstar}_R]^P_{rr}(\varepsilon_{q}^0+\hbar\omega_{{\bf k}})\big\}\big\{1-[\widetilde\Pi_{(0)}^{R}]_{rq}(-\hbar\omega_{{\bf k}'})\hbar[\widetilde{\Sigma}^{\bigstar}_R]_{rr}^P(\varepsilon_{q}^0-\hbar\omega_{{\bf k}'})\big\}}.
\end{split}
\end{equation} 
\end{widetext}
The numerator is of the form $[{1}+\widetilde{\alpha}_{\textrm{RPA}}{\Lambda}]^2,$ while, in light of Eq. (\ref{den}), the denominator is of the form $[{1}-\widetilde{\alpha}_M{\Lambda}\widetilde{\alpha}_{\textrm{RPA}}{\Lambda}]^2;$ both are perfect squares when $\omega_{{\bf k}'}=\omega_{\bf k}.$ Further, with Eq. (\ref{den}), the following incident and Raman {(anti-)Stokes} shifted (unitless) enhancement factors \cite{Kerker1980,Kerker1984} may be rigorously defined by
\begin{equation}
\label{gfac}
\begin{split}
g_{rq}^{\prime\xi\gamma}(-\hbar\omega_{{\bf k}'})&=\frac{\delta^{\xi\gamma}+\widetilde{\alpha}_{\textrm{RPA}}^{\xi\delta}(-\omega_{{\bf k}'}){\Lambda}^{\delta\gamma}}{1-[\widetilde\Pi_{(0)}^{R}]_{rq}(-\hbar\omega_{{\bf k}'})\hbar[\widetilde{\Sigma}^{\bigstar}_R]^P_{rr}(\varepsilon_{q}^0-\hbar\omega_{{\bf k}'})}\\
g_{rq}^{\eta\sigma}(\hbar\omega_{\bf k})&=\frac{\delta^{\eta\sigma}+{\Lambda}^{\eta\beta}\widetilde{\alpha}_{\textrm{RPA}}^{\beta\sigma}(\omega_{\bf k})}{1-[\widetilde\Pi_{(0)}^{R}]_{rq}(\hbar\omega_{\bf k})\hbar[\widetilde{\Sigma}^{\bigstar}_R]_{rr}^P(\varepsilon_{q}^0+\hbar\omega_{{\bf k}})}
\end{split}
\end{equation}
in terms of which the transition amplitude becomes
\begin{equation}
\label{serstmat}
\begin{split}
\langle\nu^{\prime}_{J}|&{\cal T}^{\textrm{int}(2)}_{fi}|\nu_{J}\rangle=-\sum_{r;J}\widetilde{E}_{0\xi}^{(-)}g_{rq}^{\prime\xi\gamma}(-\hbar\omega_{{\bf k}'})\\
&\times\langle\nu^{\prime}_{J}|[\widetilde{\alpha}_M]^{\gamma\eta}_{pq,rr}(\hbar\omega_{\bf k},-\hbar\omega_{{\bf k}'})|\nu_{J}\rangle g_{rq}^{\eta\sigma}(\hbar\omega_{\bf k})\widetilde{E}_{0\sigma}^{(+)}.
\end{split}
\end{equation}
Due to this factorization, the quantum-mechanical Raman-scattering intensity associated with the interacting molecule-particle system displays a fourth-power enhancement when both incident and Raman scattered fields share the same frequency, in analogy to the classical case of two coupled dipoles discussed in Sec. \ref{CSERS}; see, {\it e.g.}, Eq. (\ref{enh}). Otherwise, ${\bm g}$ and ${\bm g}'$ do not maximally constructively multiply but, rather, contribute a factor of $|gg'|^2$ to the enhanced Raman-scattering intensity.

Upon expanding the molecular-electronic polarizability $\widetilde{\bm\alpha}_M$ in Eq. (\ref{serstmat}) around the molecule's equilibrium geometry ${\bf Q}_0$ and keeping only the linear term, subsequent application of Fermi's golden rule yields the following quantum-mechanical result for the SERS intensity 
\begin{widetext}
\begin{equation}
\label{IQSERS}
\begin{split}
\frac{I_{\textrm{SERS}}({\bf k}')}{I_0({\bf k})}=\frac{\omega_{\bf k}\omega^3_{{\bf k}'}}{c^4}({\cal N}'+1&)_{{\bf k}'\lambda'}\Big|\sum_{r;J}\hat{\epsilon}^{(-)\xi}_{\lambda'}({\bf k}')\langle\nu^{\prime}_{J}|({\bf Q}_J-{\bf Q}_{0})\cdot\nabla_{{\bf Q}_J}[\widetilde{\alpha}_M]^{\gamma\eta}_{pq,rr}(\hbar\omega_{\bf k},-\hbar\omega_{{\bf k}'})|\nu_{J}\rangle\hat{\epsilon}^{(+)\sigma}_{\lambda}({\bf k})\\
\times&\frac{\delta^{\xi\gamma}\delta^{\eta\sigma}+\widetilde{\alpha}_{\textrm{RPA}}^{\xi\delta}(-\omega_{{\bf k}'}){\Lambda}^{\delta\gamma}\delta^{\eta\sigma}+\delta^{\xi\gamma}{\Lambda}^{\eta\beta}\widetilde{\alpha}_{\textrm{RPA}}^{\beta\sigma}(\omega_{\bf k})+\widetilde{\alpha}_{\textrm{RPA}}^{\xi\delta}(-\omega_{{\bf k}'}){\Lambda}^{\delta\gamma}{\Lambda}^{\eta\beta}\widetilde{\alpha}_{\textrm{RPA}}^{\beta\sigma}(\omega_{\bf k})}{\big[1-\sum_{tu}{\textrm{Tr}}\big\{\widetilde{\bm\alpha}^{M}_{rq,ut}(\pm\omega_{{\bf k},{\bf k}'})\cdot{\bm\Lambda}\cdot{\tint}({d\omega'}/{2\pi i})\widetilde{\bm\alpha}_{\textrm{RPA}}(\omega')[\widetilde G^R_{(0)}]_{tu}(\varepsilon_{q}^0/\hbar\pm\omega_{{\bf k},{\bf k}'}+\omega')\cdot{\bm\Lambda}\big\}\big]^2}\Big|^2
%&\frac{\omega^3_{\bf k}\omega_{{\bf k}'}}{c^4}({\cal N}'+1)_{{\bf k}'\lambda'}\Big|\sum_r\hat{\epsilon}^{(-)\xi}_{\lambda'}({\bf k}')\frac{\delta^{\xi\gamma}+\widetilde{\alpha}_{\textrm{RPA}}^{\xi\delta}(-\omega_{{\bf k}'}){\Lambda}^{\delta\gamma}}{1-\sum_{tu}{\textrm{Tr}}\big\{\widetilde{\bm\alpha}^{M}_{rq',ut}(-\omega_{{\bf k}'})\cdot{\bm\Lambda}\cdot{\tint}({d\omega'}/{2\pi i})\widetilde{\bm\alpha}_{\textrm{RPA}}(\omega')\widetilde G^{(0)}_{tu}(\varepsilon_{q'}^0/\hbar-\omega_{{\bf k}'}+\omega')\cdot{\bm\Lambda}\big\}}\\
%&[\widetilde{\alpha}^M]^{\gamma\eta}_{p'q',rr}(\hbar\omega_{\bf k},-\hbar\omega_{{\bf k}'})\hat{\epsilon}^{(+)\sigma}_{\lambda}({\bf k})\frac{\delta^{\xi\gamma}\delta^{\eta\sigma}+\widetilde{\alpha}_{\textrm{RPA}}^{\xi\delta}(-\omega_{{\bf k}'}){\Lambda}^{\delta\gamma}\delta^{\eta\sigma}+\delta^{\xi\gamma}{\Lambda}^{\eta\beta}\widetilde{\alpha}_{\textrm{RPA}}^{\beta\sigma}(\omega_{\bf k})+\widetilde{\alpha}_{\textrm{RPA}}^{\xi\delta}(-\omega_{{\bf k}'}){\Lambda}^{\delta\gamma}{\Lambda}^{\eta\beta}\widetilde{\alpha}_{\textrm{RPA}}^{\beta\sigma}(\omega_{\bf k})}{\big[1-\sum_{tu}{\textrm{Tr}}\big\{\widetilde{\bm\alpha}^{M}_{rq',ut}(\omega_{\bf k})\cdot{\bm\Lambda}\cdot{\tint}({d\omega'}/{2\pi i})\widetilde{\bm\alpha}_{\textrm{RPA}}(\omega')\widetilde G^{(0)}_{tu}(\varepsilon_{q'}^0/\hbar+\omega_{\bf k}+\omega')\cdot{\bm\Lambda}\big\}\big]}\Big|^2
\end{split}
\end{equation} 
\end{widetext}
in the ${\bf k}'$ direction with polarization ${\bm\epsilon}^{(-)}_{\lambda'}({\bf k}')$ from a single molecule that is interacting with its self-induced plasma density fluctuations in a nearby classical metallic particle in the presence of an external radiation source propagating in the ${\bf k}'$ direction with polarization ${\bm\epsilon}_{\lambda}^{(+)}({\bf k}).$ To our knowledge, this is the first place in the literature where a SERS intensity has been derived entirely from first principles that explicitly treats the coupling and back-reaction effects of a quantum-mechanical molecular-electronic system with a nearby metallic particle supporting collective excitation of conduction electrons in the presence of a perturbing radiation field. Both image and local-field effects, which contain the essential physics underlying the electromagnetic mechanism of SERS, have been generalized beyond the classical model of Sec. \ref{CSERS} to a quantum-mechanical setting.

We note that it is not possible to exactly factor this general expression into a normal Raman-scattering component and an enhancement factor as was done in the classical case beyond what was already written in Eqs. (\ref{gfac}) and (\ref{serstmat}). This is due to the summation over the molecular states $r$ and contraction on spatial (Greek) indices that connect all parts of the expression together. Nonetheless, it is still possible to identify a normal Raman-like component and an enhancement factor as the prefactor and quotient respectively in Eq. (\ref{IQSERS}). As in the case of normal Raman scattering, $\langle\nu^{\prime}_{J}|({\bf Q}_J-{\bf Q}_{0})|\nu_{J}\rangle$ is nonzero whenever $\nu^{\prime}_{J}=\nu_{J}\pm1.$ Note that, due to formatting constraints, the square in the denominator is to be understood as the product of two separate factors: one associated with an incident photon with frequency $\omega_{\bf k}$ and another associated with a Raman {(anti-)Stokes} scattered photon with frequency $\omega_{{\bf k}'}.$

\subsection{\label{LRT0}Molecular and Particle Response}
In the previous sections we have developed both nonperturbative and perturbative expressions for the one-body molecular-electronic Green's function that built in the effects of interaction with a nearby metal particle under the influence of an external radiation field. Together with the scattering $T$-matrix it was demonstrated that these Green's function techniques underlie a quantum-mechanical picture of SERS that generalizes (and, in the appropriate limits, reduces to) the classical theory presented in Sec. \ref{CSERS}.

Until this point only perturbations acting upon the molecular system have been treated. Where the particle's response to the molecule and field is prescribed by the RPA polarizability (\ref{RPAa}) or some variant, {\it e.g.}, a Drude polarizability, there is no need to explicitly quantify their perturbing effects. However, it is our desire to go beyond the RPA and solve for the classical dynamics of the particle's first-order induced dipole moment ${\bf p}^{(1)}$ and resultant polarizability ${\bm\alpha}_P$ as set up by a nearby Raman-active molecule (both image and local-field effects) and external radiation field. Incorporation of the latter interaction is straightforward, however, before including the former, we must first compute the induced molecular-electronic dipole moment ${\bf d}^{(1)}$ of the interacting molecular system. Knowledge of ${\bf d}^{(1)}$ closes our theoretical formulation of the many-body SERS problem and renders it computationally well-defined. It is our goal here to compute this molecular response and develop basic governing equations for the dynamics of the metallic particle.

\subsubsection{\label{LRT}Review of linear-response theory}
The linear response of a quantum many-body system \cite{Kubo57,FW,Ring,KB62} subjected to a sufficiently weak external time-dependent perturbation $\hat{\cal V}(t)=\int\hat\rho({\bf x},t)\upsilon({\bf x},t)d^3x$ may be characterized by the first-order fluctuations in its charge density $\hat\rho$ according to
\begin{equation}
\label{resp}
\begin{split}
&\langle\hat\rho({\bf x},t)\rangle_{\textrm{ext}}-\rho({\bf x},t)=\rho^{(1)}({\bf x},t)+\cdots\\
&=\frac{i}{\hbar}{\tint}d^3x'{\tint}_{-\infty}^tdt'\langle\big[\hat\rho({\bf x}',t'),\hat\rho({\bf x},t)\big]\rangle\upsilon({\bf x}',t')+\cdots,
\end{split}
\end{equation}
where the expectation value $\langle\cdots\rangle$ is taken within some unperturbed many-particle reference state, while the many-particle states underlying $\langle\cdots\rangle_{\textrm{ext}}$ include the effects of the external perturbation $\hat{\cal V};$ $\langle\cdots\rangle_{\textrm{ext}}$ reduces to $\langle\cdots\rangle$ in the limit of $\hat{\cal V}\to0.$

For definiteness and continuity with the previous we henceforth choose the states underlying $\langle\cdots\rangle$ to be the noninteracting HF reference state $|\Phi^N_{\textrm{HF}}\rangle$ and work in the interaction picture with respect to it. From Eq. (\ref{resp}), we see that the zeroth-order (static) unperturbed HF density $\rho=\langle\Phi^N_{\textrm{HF}}|\hat\rho|\Phi^N_{\textrm{HF}}\rangle,$ while the first-order density fluctuations
\begin{equation}
\label{HFrho}
\rho^{(1)}({\bf x},t)=\frac{1}{\hbar}{\tint}d^4x'\Pi^R_{(0)}({\bf x},t;{\bf x}',t')\upsilon({\bf x}',t')
\end{equation}
can be reexpressed in terms of the retarded HF polarization propagator $\Pi^R_{(0)}$ defined in Eq. (\ref{retPi}). Here, the time integral is extended to all times through the definition
\begin{equation}
\begin{split}
i\Pi^R_{(0)}({\bf x},t;{\bf x}'&,t')=\theta(t-t')\langle\Phi^N_{\textrm{HF}}|\big[\hat\rho({\bf x},t),\hat\rho({\bf x}',t')\big]|\Phi^N_{\textrm{HF}}\rangle\\
&=\theta(t-t')\langle\Phi^N_{\textrm{HF}}|\big[\delta\hat\rho({\bf x},t),\delta\hat\rho({\bf x}',t')\big]|\Phi^N_{\textrm{HF}}\rangle,
\end{split}
\end{equation}
where, as before, $\delta\hat\rho=\hat\rho-\langle\Phi^N_{\textrm{HF}}|\hat\rho|\Phi^N_{\textrm{HF}}\rangle=\hat\rho-\rho.$

%Further, we assume here and in the following that all density fluctuations beyond first order are so small that $\hat\rho_{{\textrm{int}}}\approx\hat\rho^{(1)}.$

%The Fourier transform of the matrix elements of this expression is identical to the expression for $[\widetilde\Pi^R_{(0)}]_{pqrs}$ displayed in the first line of Eq. (\ref{retPi}). 

Taking the external perturbing potential $\upsilon({\bf x},t)=-(-e{\bf x})\cdot{\bf E}_{0}({\bf x}={\bf 0},t)$ in the electric dipole-interaction approximation, we can derive an explicit expression for the first-order induced dipole moment ${\bf d}^{(1)}$ generated by $\hat{\cal V}.$ It is
\begin{equation}
\label{kernel2}
\begin{split}
{\bf d}^{(1)}(t)&={\tint}d^3x(-e{\bf x})\rho^{(1)}({\bf x},t)\\
&=\frac{-1}{\hbar}{\tint}d^3xd^4x'(-e{\bf x})\Pi^R_{(0)}({\bf x},t;{\bf x}',t')(-e{\bf x}')\cdot{\bf E}_{0}(t')\\
&={\tint}dt'{\bm\alpha}_M(t,t')\cdot{\bf E}_{0}(t'),
\end{split}
\end{equation}
where the molecular-electronic HF polarizability (\ref{HFalph}) may also be defined by
\begin{equation}
\begin{split}
-i\hbar{\bm\alpha}_M(t,t')&={\tint}d^3xd^3x'(-e{\bf x})i\Pi^R_{(0)}({\bf x},t;{\bf x}',t')(-e{\bf x}')\\
%&={\tint}d^3xd^3x'(-e{\bf x})\theta(t-t')\\
%&\ \ \ \times\langle\Phi^N_{\textrm{HF}}|\big[\hat\rho_{{\textrm{int}}}({\bf x},t),\hat\rho_{{\textrm{int}}}({\bf x}',t')\big]|\Phi^N_{\textrm{HF}}\rangle(-e{\bf x}')\\
&=\theta(t-t')\langle\Phi^N_{\textrm{HF}}|\big[\delta\hat{\bf d}(t),\delta\hat{\bf d}(t')\big]|\Phi^N_{\textrm{HF}}\rangle.
\end{split}
\end{equation}
From Eqs. (\ref{kernel2}) and (\ref{retPi}), we see that, through the underlying polarization propagator $\Pi^R_{(0)},$ the polarizability ${\bm\alpha}_M$ acts as a integral response kernel that depends upon the set of all single-particle electronic states of the molecule and parametrically upon all underlying nuclear coordinates. The external field ${\bf E}_0$ induces electronic density fluctuations that oscillate with the field until the field frequency is close to an excitation energy of molecule. In this way, information on molecular electronically-excited states and transition amplitudes can be obtained, among other quantities.

\subsubsection{\label{LRT2}Molecular response to perturbations induced by a metallic particle and external electric field}
Equation (\ref{HFrho}) prescribes a method to compute the first-order density fluctuations of a noninteracting molecular system induced by the external potential $\upsilon({\bf x},t)=-(-e{\bf x})\cdot{\bf E}_{0}(t)$ from integration of $\upsilon$ against the kernel $\Pi^R_{(0)}.$ The quantity $\rho^{(1)}$ describes how the molecular-electronic density alone responds to the perturbation $\upsilon.$ However, we are not interested in knowing how the noninteracting molecular system responds to $\upsilon,$ but, rather, want to study the response of an interacting molecule-particle system. Replacing the noninteracting integral kernel $\Pi^R_{(0)}$ with the interacting polarization propagator $\Pi{^R_M}$ describing the coupling between molecular electrons and conduction electrons in the particle achieves precisely this goal. In particular, it is desired for $\Pi{^R_{M}}$ to incorporate those effects already built into the interacting molecular-electronic Green's function ${\cal G}^P.$

A simple and general relationship exists between the polarization propagator and the one-body Green's function. Application of Wick's theorem to the full $\Pi{_M},$ defined in analogy to Eq. (\ref{HFPi}), reveals that 
%\begin{widetext}
\begin{equation}
\label{PiGG}
\begin{split}
i&\Pi{_M}({\bf x},t;{\bf x}',t')\\
&=\langle\Phi^N_{\textrm{HF}}|T\{\hat\Psi^\dagger({\bf x},t)\hat\Psi({\bf x},t)\hat\Psi^\dagger({\bf x}',t')\hat\Psi({\bf x}',t')\}|\Phi^N_{\textrm{HF}}\rangle\\
&-\langle\Phi^N_{\textrm{HF}}|\hat\Psi^\dagger({\bf x},t)\hat\Psi({\bf x},t)|\Phi^N_{\textrm{HF}}\rangle\langle\Phi^N_{\textrm{HF}}|\hat\Psi^\dagger({\bf x}',t')\hat\Psi({\bf x}',t')|\Phi^N_{\textrm{HF}}\rangle\\
%&=\langle\Phi^N_{\textrm{HF}}|\hat\Psi^\dagger({\bf x},t)\hat\Psi({\bf x},t)|\Phi^N_{\textrm{HF}}\rangle\langle\Phi^N_{\textrm{HF}}|\hat\Psi^\dagger({\bf x}',t')\hat\Psi({\bf x}',t')|\Phi^N_{\textrm{HF}}\rangle-\langle\Phi^N_{\textrm{HF}}|T\{\hat\Psi({\bf x},t)\hat\Psi^\dagger({\bf x}',t')\}|\Phi^N_{\textrm{HF}}\rangle\langle\Phi^N_{\textrm{HF}}|T\{\hat\Psi({\bf x}',t')\hat\Psi^\dagger({\bf x},t)\}|\Phi^N_{\textrm{HF}}\rangle\\
%&\ \ \ -\langle\Phi^N_{\textrm{HF}}|\hat\Psi^\dagger({\bf x},t)\hat\Psi({\bf x},t)|\Phi^N_{\textrm{HF}}\rangle\langle\Phi^N_{\textrm{HF}}|\hat\Psi^\dagger({\bf x}',t')\hat\Psi({\bf x}',t')|\Phi^N_{\textrm{HF}}\rangle\\
&=-i{\cal G}^P({\bf x},t;{\bf x}',t')i{\cal G}^P({\bf x}',t';{\bf x},t),
\end{split}
\end{equation}
%\end{widetext}
where the electron field operators are expanded onto the underlying basis of interacting orbitals $\chi_q$ and interacting electronic creation and annihilation operators $\hat C_q$ defined in Sec. \ref{mep}. This result together with Eq. (\ref{HFrho}), demonstrates that the first-order density fluctuations of the interacting molecule-particle system may be computed from 
\begin{equation}
\begin{split}
\rho^{(1)}_{M}({\bf x},t)&=\frac{1}{\hbar}{\tint}d^4x'\Pi{^R_{M}}({\bf x},t;{\bf x}',t')\upsilon({\bf x}',t')\\
&=\frac{1}{i\hbar}{\tint}d^4x'\big[{\cal G}^P({\bf x},t;{\bf x}',t'){\cal G}^P({\bf x}',t';{\bf x},t)\big]_R\upsilon({\bf x}',t'),
\end{split}
\end{equation}
or, alternatively, from the spectral form
\begin{equation}
\begin{split}
&\widetilde\rho^{(1)}_{M}({\bf x},\omega)=\frac{1}{\hbar}{\tint}d^3x'\widetilde\Pi{^R_{M}}({\bf x},{\bf x}';\omega)\widetilde\upsilon({\bf x}',\omega)\\
&=\frac{1}{\hbar}{\tint}d^3x'\Big[{\tint}\frac{d\omega'}{2\pi i}\widetilde{\cal G}^P({\bf x},{\bf x}';\omega+\omega')\widetilde{\cal G}^P({\bf x}',{\bf x};\omega')\Big]_R\widetilde\upsilon({\bf x}',\omega)
\end{split}
\end{equation}
with $pq$-matrix elements
\begin{equation}
\label{pirmat}
\begin{split}
[\widetilde\rho^{(1)}_{M}]_{pq}(\omega)&=\frac{1}{\hbar}\sum_{rs}[\widetilde\Pi{^R_{M}}]_{pqrs}(\omega)\widetilde\upsilon_{sq}(\omega)\\
&=\frac{1}{\hbar}\sum_{rs}\Big[{\tint}\frac{d\omega'}{2\pi i}\widetilde{\cal G}^P_{pr}(\omega+\omega')\widetilde{\cal G}^P_{sq}(\omega')\Big]_R\widetilde\upsilon_{sq}(\omega).
\end{split}
\end{equation}
From this last expression it is clear that knowledge of $\widetilde\rho^{(1)}_{M}$ may be attained once the (retarded component of the) interacting molecular polarization propagator $\widetilde\Pi{_{M}}$ is known. By assuming that the self energy varies so slowly with frequency that it is appropriate to approximate $[\widetilde{\Sigma}^{\bigstar}_P]_{pp}(\omega)\approx[\widetilde{\Sigma}^{\bigstar}_P]_{pp}(\varepsilon_p^0/\hbar)$ and $[\widetilde{\Sigma}^{\bigstar}_P]_{pq}(\omega)\approx[\widetilde{\Sigma}^{\bigstar}_P]_{pq}(\varepsilon_p^0/\hbar),$ the interacting $\widetilde\Pi{^R_{M}}$ reduces to
\begin{equation}
\begin{split}
&[\widetilde\Pi{^R_{M}}]_{pqrs}(\omega)=\frac{\delta_{pr}\delta_{sq}}{\omega-\varepsilon_p^0/\hbar+\varepsilon_s^0/\hbar-[\widetilde{\Sigma}_R^{\bigstar}]^P_{pp}+[\widetilde{\Sigma}_A^{\bigstar}]^P_{qq}}\\
&=\frac{\delta_{pr}\delta_{sq}}{\omega-\varepsilon_p^0/\hbar+\varepsilon_s^0/\hbar-[\Delta^P_{pp}-(i/2)\Gamma^P_{pp}]+[\Delta^P_{qq}+(i/2)\Gamma^P_{qq}]}
\end{split}
\end{equation}
in the weak-coupling limit. The shifting ($\Delta^P$) and broadening ($\Gamma^P$) of the single-particle states $p$ and $s$ due to the explicit treatment of interaction have been quantified previously in Eqs. (\ref{shift}) and (\ref{broad}).

Together with Eq. (\ref{pirmat}), this approximate expression for the interacting molecular-electronic polarization propagator can be used to compute the first-order electronic density fluctuations $\widetilde\rho^{(1)}_M$ in the molecule induced by the external field ${\bf E}_0$ and, through $\widetilde{\cal G}^P,$ by interaction with molecule-induced density fluctuations $n_M$ in a nearby classical metal particle. In the dipole approximation, the electric field at the point ${\bf x}_2$ generated from $\widetilde\rho^{(1)}_M$ located at the origin is given by
\begin{equation}
\widetilde{\bf E}_M({\bf x}_2,\omega)={\bm\Lambda}\cdot\widetilde{\bf d}^{(1)}_M(\omega)={\bm\Lambda}\cdot{\tint}(-e{\bf x})\widetilde\rho^{(1)}_M({\bf x},\omega)d^3x
\end{equation}
in the frequency domain. It will be now be demonstrated how the induced local molecular electric field $\widetilde{\bf E}_M$ influences the dynamics of the conduction electrons in a nearby particle.

\subsubsection{\label{PRP1}Particle response to perturbations induced by molecular electrons and external electric field}
In order to expose certain notable features of our formalism, such as the enhanced Raman-scattering intensity in Eq. (\ref{IQSERS}), we have imposed a predetermined dynamics upon the collective excitations of conduction electrons in the metal particle: namely, that of the RPA high-density electron gas (\ref{RPAa}). This {\it ad hoc} choice has facilitated analytical derivation. Here we provide the minimal theoretical framework necessary to go beyond such a formalism and treat the dynamics of the metal particle's conduction electrons explicitly.

In Eq. (\ref{partH}) of Sec. \ref{QSERS}, the many-body Hamiltonian $\hat H_P$ of the particle and its interaction with a nearby Raman-active molecule under the influence of an external electric field was presented. Its external interaction component can be decomposed according to
\begin{equation}
%\begin{split}
\hat H^P_{\textrm{int}}(t)\approx-{\bf p}(t)\cdot[{\bf E}_{0}({\bf x}_2,t)+{\bm\Lambda}\cdot\hat{\bf d}^{(1)}_M(t)],
%\end{split}
\end{equation}
where the interaction potential $W\approx-{\bf p}\cdot{\bm\Lambda}\cdot{\bf d}^{(1)}_M$ has been multipole expanded and only the term of dipole order was retained. Linear-response theory may now be employed to compute the electronic density fluctuations induced in the particle by interactions with the external field ${\bf E}_0$ and with the dipole field of the molecule ${\bm\Lambda}\cdot{\bf d}^{(1)}_M.$ This latter term represents the local-field effect of the molecule upon the particle.

By discretizing the particle's associated electric dipole fluctuations onto a three-dimensional spatial grid, the resulting response equation for the $j$th induced dipole moment is given by
\begin{equation}
\label{discretea}
\begin{split}
%{\bf p}_{\textrm{int}}(t)&=\frac{1}{\hbar}{\tint}d^3xd^4x'(-e{\bf x})\Pi{_P}({\bf x},t;{\bf x}',t')(-e{\bf x}')\\
%&\ \ \ \cdot[{\bf E}_0(t')+\frac{1}{2}{\bm\Lambda}\cdot{\bf p}(t')+{\bm\Lambda}\cdot\hat{\bf d}_{\textrm{int}}^M(t')]\\
{\bf p}^{(1)}_j(t)={\tint}&dt'{\bm\alpha}_{Pj}(t,t')\\
&\cdot[{\bf E}_{0j}(t')+\sum_{k\neq j}{\bm\Lambda}'_{jk}\cdot{\bf p}^{(1)}_k(t')+{\bm\Lambda}_{j}\cdot{\bf d}^{(1)}_{M}(t')],
\end{split}%${\bf E}_{0j}(t)={\bf E}_{0}^{(+)}({\bf x}_{2j})\exp(-i\omega_{\bf k}t)+{\bf E}_{0}^{(-)}({\bf x}_{2j})\exp(i\omega_{\bf k}t),$
\end{equation}
where ${\Lambda}^{\prime\xi\sigma}_{jk}=[3\hat r_{jk}^\xi\hat r_{jk}^\sigma-\delta^{\xi\sigma}]/r_{jk}^3$ with $r_{jk}=|{\bf x}_{2j}-{\bf x}_{2k}|$ is the dipole tensor associated with the interaction of two such discretized induced dipoles in the particle labeled by $j,k$ ($k\neq j$), and ${\Lambda}^{\xi\sigma}_{j}=[3\hat r_{j}^\xi\hat r_{j}^\sigma-\delta^{\xi\sigma}]/r_{j}^3$ with $r_{j}=|{\bf x}_{1}-{\bf x}_{2j}|$ is the dipole tensor associated with the interaction of the discretized induced dipole in the particle labeled by $j$ with the induced molecular dipole moment $\hat{\bf d}^{(1)}_M$ located at ${\bf x}_{1},$ which may be taken as the coordinate origin. The underlying particle polarizability ${\bm\alpha}_{Pj}$ in Eq. (\ref{discretea}) may be approximated by an appropriately discretized polarizability that is consistent with the optical theorem \cite{draine94}. Both image and local-field effects are included in this expression by the terms $\sum_{k\neq j}{\bm\Lambda}'_{jk}\cdot{\bf p}^{(1)}_k$ and ${\bm\Lambda}_{j}\cdot{\bf d}^{(1)}_{M}$ respectively.

In Fourier space, ${\bf p}^{(1)}_j$ takes the form
\begin{equation}
%\begin{split}
\widetilde{\bf p}^{(1)}_j(\omega)=\widetilde{\bm\alpha}_{Pj}(\omega)\cdot[\widetilde{\bf E}_{0j}(\omega)+\sum_{k\neq j}{\bm\Lambda}'_{jk}\cdot\widetilde{\bf p}_{k}^{(1)}(\omega)+{\bm\Lambda}_j\cdot\widetilde{\bf d}^{(1)}_M(\omega)],
%\end{split}
\end{equation}
which may be inverted to yield
\begin{equation}
\sum_{k}\big[\widetilde{\bm\alpha}_{Pj}^{-1}(\omega)\delta_{jk}-(1-\delta_{jk}){\bm\Lambda}'_{jk}\big]\widetilde{\bf p}_k^{(1)}(\omega)=\widetilde{\bf E}_{0j}(\omega)+\widetilde{\bf E}_{Mj}(\omega).
\end{equation}
Hence, discretization leads to a linear $AX=B$ system of equations that are solvable by standard linear algebra routines \cite{Lapack}. Similar response equations are well known in the literature; see, {\it e.g.}, Ref. \cite{draine94}. This spatial discretization enables the practical treatment of anisotropic metal particles of arbitrary shape and size.

In the dipole approximation, the (continuum) electric field stemming from the solution of these equations may be written as $\widetilde{\bf E}_P={\bm\Lambda}\cdot\widetilde{\bf p}^{(1)}.$ By allowing the induced electric field of the particle's conduction electrons to act back upon the molecular system through, {\it e.g.}, Eq. (\ref{fep1}), the many-body formalism presented in this article is mathematically closed and well defined. Iteration between a quantum-chemical molecular response calculation where the incident field $\widetilde{\bf E}_0$ and induced field of the particle $\widetilde{\bf E}_P$ enter as perturbations and a classical-electrodynamical metal particle response calculation where the incident field $\widetilde{\bf E}_0$ and induced field of the molecule $\widetilde{\bf E}_M$ enter as perturbations should be performed until self consistency is reached between the two systems.

\section{Summary and Conclusions}
We have presented a unified and didactic approach to the understanding of the microscopic theory of single-molecule SERS from a nanoscale metal particle at zero temperature. Nonperturbative and perturbative many-body Green's function techniques are employed to build in the interaction between a molecular-electronic system and the conduction electrons of a nearby metallic particle in the presence of an external radiation field from first principles. Both image and local-field effects between molecule and metal are explicitly included. Due to its generality, other relevant approaches from the literature, including those that are purely classical and purely quantum-mechanical, are obtained by taking appropriate limits of our formalism.

With emphasis placed upon practical initial numerical implementation, molecular-electronic correlation is restricted to the level of Hartree-Fock mean-field theory, while a Bogoliubov decomposition of the metallic particle's plasmon field is effected to reduce these collective electronic excitations to a classical field; extension of former through a Kohn-Sham density-functional or M{\o}ller-Plesset perturbation theory is discussed; specialization of the latter to the RPA allows for the analytic presentation of several salient features of the theory such as the enhanced Raman-scattering intensity in Eq. (\ref{IQSERS}) and incident and Raman {(anti-)Stokes} scattered enhancement factors in Eq. (\ref{gfac}). We believe that such expressions, where interaction effects are systematically included from first-principles, have never before appeared in the literature.

Going beyond the RPA, we explicitly describe the response of the metal particle's conduction electrons to the external field and to the induced electronic density in a nearby molecule by discretizing the associated classical response equations onto a spatial grid. By iteration until self consistency is reached, the reaction and back-reaction effects between molecular and particle systems with each other and with the external perturbing radiation field may be incorporated: thereby mathematically closing the theory. Having laid out our many-body formalism in this article and demonstrated its relation to and generalization of other approaches from the literature, as well as some of its important properties, implementation is currently underway to numerically realize these equations and provide theoretical support to current single-molecule SERS experiments.

\appendix
\section{\label{appDFT}Inclusion of electron-electron interaction effects}
The effects of electron-electron interaction stemming from $\hat H_{\textrm{int}}$ in Eq. (\ref{hint}) may be straightforwardly included in the above formalism with either M{\o}ller-Plesset perturbation theory or density-functional theory. In the latter case, the HF orbital equation (\ref{HFE}) is replaced by the Kohn-Sham orbital equation 
\begin{equation}
\begin{split}
&0=[h_0({\bf x})+U_0({\bf x})-\varepsilon^{\textrm{KS}}_q]\varphi_q({\bf x})\\
&\ \ \ +\sum_k\langle k|V|k\rangle({\bf x})\varphi_q({\bf x})+\big(\delta E^M_{\textrm{XC}}[\rho]/\delta\rho({\bf x})\big)\varphi_q({\bf x}),
\end{split}
\end{equation}
where the total molecular-electronic energy 
\begin{equation}
\begin{split}
E_M&={\tint}\rho({\bf x})[U_0({\bf x})+V_{\textrm{ext}}({\bf x})]d^3x\\
&\ \ \ +(1/2){\tint}\rho({\bf x})V({\bf x},{\bf x}')\rho({\bf x}')d^3xd^3x'+G[\rho]
\end{split}
\end{equation}
may be expressed as the sum of the external electron-nuclear attraction $V_{\textrm{ext}}$ and universal functional
\begin{equation}
G[\rho]=T_0[\rho]+E^M_{\textrm{XC}}[\rho]
\end{equation}
with noninteracting electronic kinetic energy $T_0\equiv T_0[\rho].$ If, in addition, $E^M_{\textrm{XC}}$ included the interaction effects among conduction electrons within the metal particle (or a subset thereof) and between these conduction electrons and the molecular-electronic system, then the present approach (with a few minor modifications) would recover the work presented in Refs. \cite{Jensen2006a,Aikens2006a,Jensen2007b,Aikens2008}. In this way, the chemical mechanism of SERS can additionally be incorporated within the present formalism.

\section{\label{PP}Polarization propagator and linear polarizability}
The noninteracting HF polarization propagator \cite{LO,FW} is defined by the expectation
\begin{widetext}
\begin{equation}
\label{HFPi}
\begin{split}
i\Pi^{(0)}({\bf x},t;{\bf x}',t')&=\langle\Phi^N_{\textrm{HF}}|T\{\delta\hat\rho({\bf x},t)\delta\hat\rho({\bf x}',t')\}|\Phi^N_{\textrm{HF}}\rangle\\
&=\sum_{pqrs}\phi^*_q({\bf x})\phi_p({\bf x})i\Pi^{(0)}_{pqrs}(t,t')\phi^*_r({\bf x}')\phi_s({\bf x}')\\
&={\tint}\frac{d\omega}{2\pi}e^{-i\omega(t-t')}\sum_{pqrs}\phi^*_q({\bf x})\phi_p({\bf x})i\Bigl\{\delta_{pr}\delta_{sq}\Bigl[\frac{(1-\rho_p^0)\rho^0_q}{\omega+i0^++(\varepsilon_q^0-\varepsilon_p^0)/\hbar}-\frac{\rho^0_p(1-\rho^0_q)}{\omega-i0^++(\varepsilon_q^0-\varepsilon_p^0)/\hbar}\Bigr]\Bigr\}\phi^*_r({\bf x}')\phi_s({\bf x}').
\end{split}
\end{equation}
\end{widetext}
It has the retarded component
\begin{widetext}
\begin{equation}
\label{retPi}
\begin{split}
i\widetilde\Pi_{(0)}^R({\bf x},t;{\bf x}',t')&=\theta(t-t')\langle\Phi^N_{\textrm{HF}}|\big[\delta\hat\rho({\bf x},t),\delta\hat\rho({\bf x}',t')\big]|\Phi^N_{\textrm{HF}}\rangle\\
&=\sum_{pqrs}\phi^*_q({\bf x})\phi_p({\bf x})i[\Pi^R_{(0)}]_{pqrs}(t,t')\phi^*_r({\bf x}')\phi_s({\bf x}')\\
&={\tint}\frac{d\omega}{2\pi}e^{-i\omega(t-t')}\sum_{pqrs}\phi^*_q({\bf x})\phi_p({\bf x})i\Big\{\delta_{pr}\delta_{sq}\frac{\rho^0_q-\rho^0_p}{\omega+i0^++(\varepsilon_q^0-\varepsilon_p^0)/\hbar}\Big\}\phi^*_r({\bf x}')\phi_s({\bf x}').
%&={\tint}\frac{d\omega}{2\pi}e^{-i\omega(t-t')}\sum_{pqrs}\phi^*_q({\bf x})\phi_p({\bf x})\delta_{pr}\delta_{sq}(\rho^0_q-\rho^0_p)i[\widetilde\Pi_{(0)}^R]_{pq}(\omega)\phi^*_r({\bf x}')\phi_s({\bf x}')
\end{split}
\end{equation}
\end{widetext}
expressed in terms of the commutator $[\cdot,\cdot].$ It represents the probability amplitude that a molecular-electronic density disturbance $\delta\hat\rho$ originating at the space-time point $({\bf x}',t')$ will be found later within the $N$-electron Fermi vacuum $|\Phi^N_{\textrm{HF}}\rangle$ at the space-time point $({\bf x},t),$ where the electron density operator $\hat\rho({\bf x},t)=\hat\Psi^\dagger({\bf x},t)\hat\Psi({\bf x},t)$ is decomposed into static and interaction-induced excitation parts according to $\hat\rho=\langle\Phi^N_{\textrm{HF}}|\hat\rho|\Phi^N_{\textrm{HF}}\rangle+\delta\hat\rho.$ Proof of Eq. (\ref{piM}) may be demonstrated by retaining the plasmon field operators $\hat\Omega$ to second order in perturbation theory in Eq. (\ref{link}), defining plasmon Green's functions, and then effecting the Bogoliubov decomposition (\ref{bogo}) together with the Heaviside identity $\theta(x)+\theta(-x)=1.$

From Eq. (\ref{retPi}), the linear HF polarizability is defined by
\begin{equation}
\label{HFalph}
\begin{split}
-&i\hbar\alpha_M^{\xi\sigma}(t,t')=\sum_{pqrs}\langle q|-ex^\xi|p\rangle i[\Pi^R_{(0)}]_{pqrs}(t,t')\langle r|-ex^\sigma|s\rangle\\
&=\theta(t-t')\langle\Phi^N_{\textrm{HF}}|\big[\delta\hat d^\xi(t),\delta\hat d^\sigma(t')\big]|\Phi^N_{\textrm{HF}}\rangle\\
&={\tint}\frac{d\omega}{2\pi i}e^{-i\omega(t-t')}\sum_{pq}\rho^0_p\Big[\frac{\langle q|-ex^\xi|p\rangle\langle p|-ex^{\prime\sigma}|q\rangle}{(\varepsilon_q^0-\varepsilon_p^0)/\hbar+\omega+i0^+}\\
&\hspace{3.75cm}+\frac{\langle q|-ex^{\prime\sigma}|p\rangle\langle p|-ex^{\xi}|q\rangle}{(\varepsilon_q^0-\varepsilon_p^0)/\hbar-\omega-i0^+}\Big]
\end{split}
\end{equation} 
with interaction-induced dipole moment $\delta\hat{\bf d}(t)=\int(-e{\bf x})\delta\hat\rho({\bf x},t)d^3x$ being related to its expectation value in an externally perturbed reference state $\langle\cdots\rangle_{\textrm{ext}}$ by $\langle\delta\hat{\bf d}(t)\rangle_{\textrm{ext}}\equiv{\bf d}^{(1)}(t)+\cdots={\tint}dt'{\bm\alpha}_M(t,t')\cdot{\bf E}_{0}(t')+\cdots;$ see Sec. \ref{LRT} for a review of linear-response theory, and, in particular Eq. (\ref{kernel2}). It can be shown that ${\bm\alpha}_M$ satisfies $[\widetilde{\bm\alpha}_M(\omega)]^*=\widetilde{\bm\alpha}_M(-\omega).$ We note that there has been an ongoing debate in the literature regarding the damping sign ($\pm i0^+$) convention in the linear (Kramers-Heisenberg) polarizability. This issue has recently been positively resolved by Bialynicki-Birula and Sowi{\'n}ski \cite{Bialynicki-Birula2007} for linear processes and by Mukamel \cite{Mukamel2007} for nonlinear processes, and the differences between scattering and response points of view have been clarified. To briefly summarize, both signs are correct but apply to different physical situations \cite{Bialynicki-Birula2007}. The fundamental principle of causality is always respected, however, response theory describes interaction processes occurring with a particular causal time ordering where the initial state and form of interaction are specified and a sum over final states is performed, while scattering theory equally includes all possible time orderings of interactions between specific initial and final states and, in this sense, is noncausal \cite{Mukamel2007}. The response functions [{\it e.g.}, Eq. (\ref{HFalph})] and transition amplitudes [{\it e.g.}, Eq. (\ref{transalpha})] computed here from a many-body perspective are consistent with Refs. \cite{Bialynicki-Birula2007,Mukamel2007}; the differences in damping sign between Eq. (\ref{HFalph}) and Eq. (\ref{transalpha}) below are correct.

In order to compute Raman transition amplitudes, we introduce the Lehmann representation of the generalized noninteracting molecular transition polarizability 
\begin{equation}
\label{transalpha}
\begin{split}
-\hbar[\widetilde\alpha_M&]^{\xi\sigma}_{pq}(\omega,-\omega')\\
&=\sum_{rs}\Big[\langle p|-ex^\xi|r\rangle\delta_{rs}[\widetilde\Pi^R_{(0)}]_{rq}(\omega)\langle s|-ex^{\prime\sigma}|q\rangle\\
&\hspace{0.9cm}+\langle p|-ex^{\prime\sigma}|r\rangle\delta_{rs}[\widetilde\Pi^R_{(0)}]_{rq}(-\omega')\langle s|-ex^\xi|q\rangle\Big]\\
&=\sum_{rs}\Big[\frac{\langle p|-ex^\xi|r\rangle\delta_{rs}\langle s|-ex^{\prime\sigma}|q\rangle}{(\varepsilon_q^0-\varepsilon_r^0)/\hbar+\omega+i0^+}\\
&\hspace{0.9cm}+\frac{\langle p|-ex^{\prime\sigma}|r\rangle\delta_{rs}\langle s|-ex^{\xi}|q\rangle}{(\varepsilon_q^0-\varepsilon_r^0)/\hbar-\omega'+i0^+}\Big]
\end{split}
\end{equation} 
between the single-particle states $q$ and $p$ with two potentially different frequencies $\omega$ and $\omega',$ defined in terms of the $rq$-matrix elements of the retarded HF polarization propagator as
\begin{equation}
[\widetilde\Pi^R_{(0)}]_{rqps}(\omega)=\delta_{rp}\delta_{sq}(\rho^0_q-\rho^0_r)[\widetilde\Pi_{(0)}^R]_{rq}(\omega).
\end{equation}
We point out that the sums in all previous expressions for the polarization propagator and polarizability extend over the set of all single-particle states of the molecule \cite{heis}.

\begin{acknowledgments}
The authors gratefully acknowledge financial support from the Department of Energy grant No. DEFG 02-03-ER15487 and the DTRA JSTO Program FA9550-06-1-0558. Further, D.M. wishes to thank Dr. Thorsten Hansen of Northwestern University for stimulating discussions.
\end{acknowledgments}

\bibliography{jila,thesis,mal,ref}

\begin{thebibliography}{67}
\bibitem[{sta()}]{star}
\bibinfo{note}{Irreducible self energies, indicated by a star ($\bigstar$), are
  self energies that cannot be separated into two pieces by cutting a
  single-particle line.}
\bibitem[{hei()}]{heis}
\bibinfo{note}{Note that the standard Kramers-Heisenberg expression for the
  linear polarizability involves summation over all $N$-body
  electronically-excited states of the molecule excluding the $N$-body ground
  state.}
\end{thebibliography}
\end{document}